# Precursor-Dependent Energetics as a Predictive Principle for Polymorph Selection in Thin Films

Hyeon Woo Kim[1,2], Han Uk Lee[1,3], Rohan Mishra[2*] & Sung Beom Cho[1,3,4*]

[1]Department of Material Science and Engineering, Ajou University, Suwon, Gyeonggi-do 16499, Republic of Korea

[2]Department of Mechanical Engineering & Material Science, Washington University in St. Louis, St. Louis, Missouri 63130, United States

[3]Department of Energy Systems Research, Ajou University, Suwon, Gyeonggi-do 16499, Republic of Korea

[4]School of Advanced Materials Science and Engineering, Sungkyunkwan University, Suwon, Gyeonggi-do 16419, Republic of Korea

*Corresponding authors.

Email: R. Mishra: rmishra@wustl.edu , S.B. Cho: csb@ajou.ac.kr

## Abstract

Vapor deposition allows for the synthesis of metastable polymorphs with unique properties, yet polymorph selection remains largely empirical due to the lack of predictive guidelines bridging thermodynamics, kinetics, and synthesis conditions. Here, we show that precursor chemistry can control metastable polymorph selection by modulating the reaction driving force governing nucleation. By integrating first-principles reaction energetics and substrate interactions into classical nucleation theory, we establish a quantitative framework that connects precursor-dependent reaction energetics to polymorph accessibility during vapor deposition. Using $Ga_2O_3$ as a model system, we demonstrate that highly reactive precursors with large reaction driving forces kinetically stabilize the metastable $\alpha$ phase, whereas low-driving-force precursors permit thermodynamic relaxation to the stable $\beta$ phase. Furthermore, precursor flow rates amplify supersaturation, expanding the kinetic window for stabilizing the elusive $\kappa$ phase. The predictive capability of this approach is further validated in the $TiO_2$ system, where precursor-dependent reaction energetics correctly capture the competitive nucleation between rutile and anatase. These results establish precursor chemistry as a tunable chemical lever for controlling nucleation kinetics and provide a predictive design principle for metastable polymorph synthesis in vapor deposition.

## Introduction

Metastable polymorphs can exhibit technologically valuable properties that differ markedly from those of the thermodynamic ground state, including ferroelectricity at sub-nanometer length scales[1], enhanced catalytic activity[2], and superhard phases[3] (e.g., diamond relative to graphite). Despite this promise, the pathways by which metastable phases form remain poorly understood, and many are still discovered serendipitously.[4-7] In practice, precursor chemistry[8, 9], rapid thermal processing[10, 11], and off-stoichiometric growth[12, 13] can bias synthesis toward metastable products. However, without predictive rules, the growth of metastable phases remains largely empirical[14]. Gallium oxide offers a vivid example: even in homoepitaxy, $\beta$-$Ga_2O_3$ can adopt unexpected orientations, such as (001) and (−201), that are not epitaxially aligned with the substrate. Moreover, in heteroepitaxy on sapphire substrates ($\alpha$-$Al_2O_3$), the thermodynamically stable $\beta$ phase frequently outcompetes metastable $\alpha$-$Ga_2O_3$ despite the substantial lattice mismatch between $\beta$-$Ga_2O_3$ and sapphire. The metastable $\kappa$-$Ga_2O_3$ phase appears even harder to predict, with a narrow process window and divergent reports of phase selection under nominally similar conditions.[15-17] Collectively, these observations point to strong competition among polymorphs separated by small differences in free energy, such that temperature, supersaturation, precursor identity, stoichiometry, and epitaxial constraints can readily tip the balance.

The stabilization of metastable phases is often linked to kinetics. A metastable phase may form first when its nucleation barrier is lower than that of the ground state, and it can persist at ambient conditions if the subsequent transformation is kinetically hindered. In principle, such barriers can be estimated within classical nucleation theory (CNT), where surface energy and driving force jointly determine the critical nucleus and the barrier height. In many systems, metastable phases are favored because they can present lower interfacial or surface energies than the ground state under the relevant formation conditions, and reported cases in $MnO_2$[18], $Al_2O_3$[19], and $ZrO_2$[20] synthesized by hydrothermal or solid-state routes are broadly consistent with this picture.

In vapor deposition, however, nucleation is typically heterogeneous. The barrier depends strongly on wetting and film–substrate interfacial energy, rather than on free-surface energies alone. Moreover, the relevant surface and interfacial energies are dependent on growth environment because surface terminations and adsorbates vary with chemical potential and flux.[21, 22] As a consequence, vapor-deposited films do not always follow trends inferred from surface energies, and a polymorph with higher clean-surface energy can still be realized; $Ga_2O_3$[23], $In_2O_3$[24], and $HfO_2$[25] provide representative examples. For example, organic precursor dependency, reactant types, catalytic gas introduction, and the modulation of these gas fluxes, along with temperature control, are key drivers of polymorph selection. This reflects additional contributions central to vapor-deposition nucleation and growth, including: (i) interface energetics and substrate interactions, (ii) epitaxial strain energy and its relaxation pathways, (iii) and reaction/attachment kinetics that can kinetically trap accessible structures under high flux or limited adatom mobility.[26] As a result, vapor deposition yields a rich diversity of metastable polymorphs, yet recipes for selecting a desired metastable phase through nucleation control remain difficult to predict.

In this work, we demonstrate that precursor chemistry can serve as a chemical lever that controls metastable polymorph selection by modulating reaction-driven nucleation kinetics. To bridge the gap between vapor-to-solid synthesis and nucleation kinetics, we reformulate CNT by mapping macroscopic process parameters onto microscopic energetic contributions. Specifically, we derive the fundamental bulk and interfacial energy terms from precursor reactivity, epitaxial strain, flow rates, and other controllable experimental variables. By situating synthesis conditions within this energy landscape, we identify the key nucleation drivers governing phase selection and validate them against existing experimental literature. We first apply this approach to $Ga_2O_3$ as a primary case study, addressing the anomalous orientation in $\beta$-homoepitaxy and the competitive nucleation between the stable $\beta$- and metastable $\alpha/\kappa$-phase in heteroepitaxy on $\alpha$-$Al_2O_3$. From these findings, we construct nucleation-based synthetic windows that visualize the selection boundaries among polymorphs. Finally, to demonstrate the generality of the framework, we extend the analysis to the $TiO_2$ system, establishing a predictive protocol for the rational design of metastable materials.

## Results and Discussion

### ***Nucleation prediction framework for vapor-phase deposition***

Figure 1a summarizes the fundamental competition between thermodynamic and kinetic factors in governing polymorphism. Along a typical reaction pathway, while the stable phase (red) possesses a larger thermodynamic driving force, the metastable phase (blue) may exhibit a significantly lower nucleation barrier. Consequently, the final crystalline product is dictated not by thermodynamic equilibrium alone, but by the relative nucleation rate ($J$).

In principle, the nucleation rates ($J$) can be calculated within the framework of CNT, expressed as:

$$J = Aexp\left[-\frac{16\pi {\gamma_{surf}}^{3}}{3n^{2}k_{B}T(\Delta G_{bulk})}\right], \quad (1)$$

where $A$ is a pre-factor, $n$ is atomic density, $k_B$ is the Boltzmann constant, $T$ is temperature in K, $\gamma_{surf}$ is the surface energy, and $\Delta G_{bulk}$ is the bulk free energy (driving force). However, directly applying this canonical formulation to vapor deposition reveals significant limitations. Conventional models often treat $\gamma_{surf}$ and $\Delta G_{bulk}$ as static, intrinsic material properties, effectively ignoring the complex physics and chemistry inherent to realistic synthesis environments. As shown in Figure 1b, vapor-phase nucleation occurs in a dynamic 'nucleation zone' where the substrate and gaseous reactants intersect. Within this zone, the energetic landscape is reshaped by both intrinsic (bulk, surface) and extrinsic (interface, epitaxial strain) contributions. Crucially, while the extrinsic contributions from the substrate are largely deterministic and fixed once the system is defined, the energetics of the gas phase can be dynamically modulated by process variables such as flow rates and temperature. This implies that the choice of precursors, and their associated chemical potentials, serve as 'kinetic handles' that can become the dominant factor in modulating nucleation kinetics and governing phase competition.

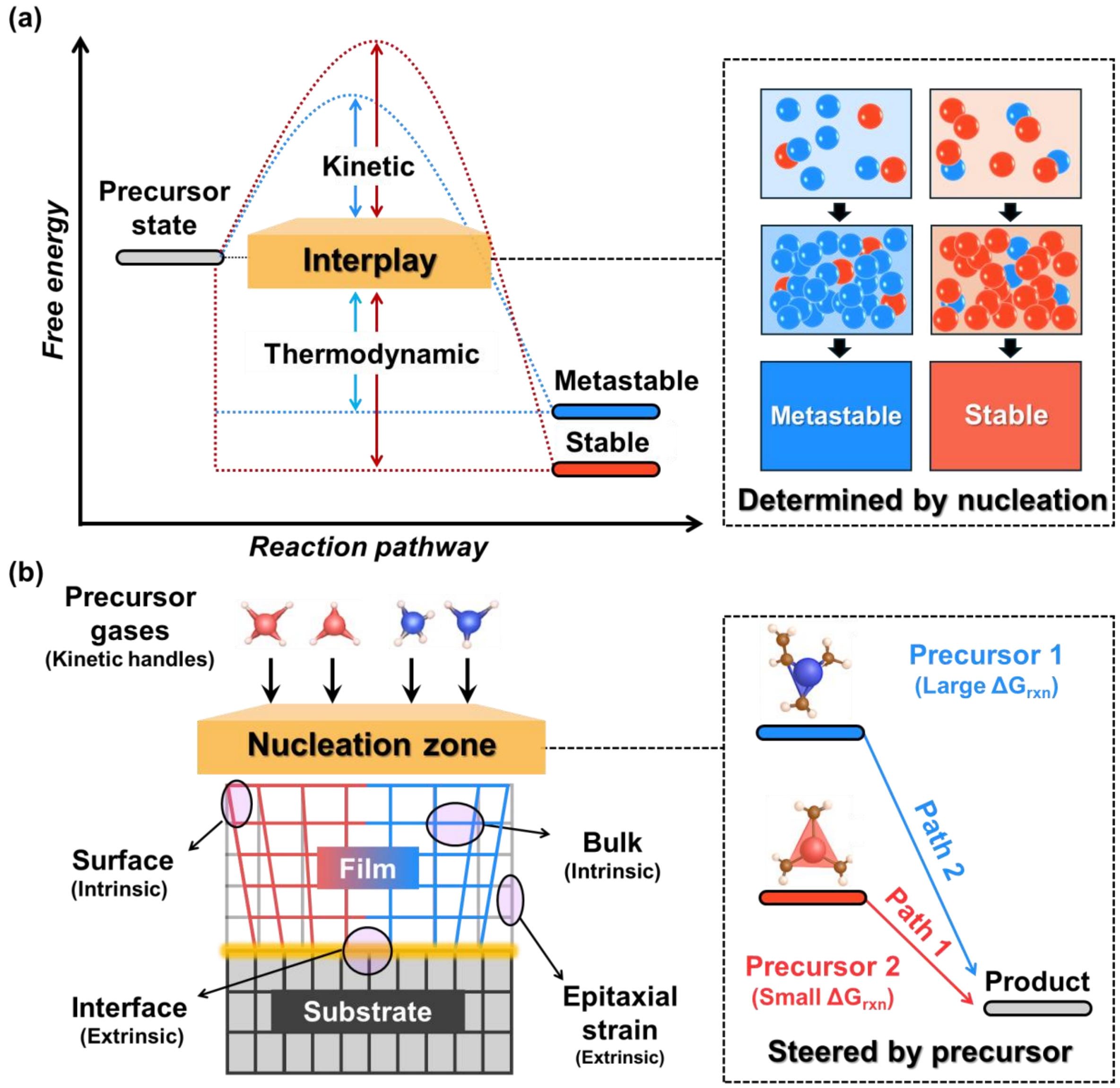

**Figure 1. Nucleation prediction logic for vapor-deposited metastable polymorphs.** (a) Free-energy landscape for crystalline materials. Phase selection is governed by the interplay between thermodynamic stability and kinetic barriers, ultimately being determined by nucleation. (b) Illustration of the nucleation zone in vapor deposition. In this zone, nucleation is kinetically steered by the precursors. The three physical domains encompass intrinsic properties (surface, bulk), extrinsic substrate effects (interface energy, epitaxial strain), and kinetic handles (precursor control). $\Delta G_{rxn}$ denotes reaction energy.

Following the discussion above, we now extend the CNT framework to the regime of vapor deposition by explicitly accounting for the interplay between intrinsic energetics, extrinsic substrate effects, and kinetic handles. This formulation highlights the reaction energy ($\Delta G_{rxn}$) as a key chemical descriptor linking precursor chemistry to nucleation kinetics. This is determined by the energy difference between the chemical potentials of the gaseous precursors and final products (See Supplementary Information Method). The term of $\Delta G_{rxn}$ includes the temperature and pressure of precursors as follows:

$$\Delta G_{rxn}(T, p_j) = \sum_{i \in products} \nu_i \mu_i^0 - \sum_{j \in precursors} \nu_j \mu_j^0 (T, p_j) \tag{2}$$

We then add the interfacial energy ($\gamma_{inter}$) and the epitaxial strain energy ($E_{epi}$) to eq. 1. Consequently, $J$ is now expressed as:

$$J = Aexp\left[-\frac{16\pi(\gamma_{surf} + \gamma_{inter})^3}{3n^2 k_B T(\Delta G_{rxn} + E_{epi})^2}\right]. \tag{3}$$

We further employ the concept of relative nucleation rates to assess phase competition, as the relative nucleation kinetics governs polymorph selection more critically than the absolute rates. By utilizing a logarithmic ratio, we defined the relative form of nucleation rates as $\ln\left(\frac{J_a}{J_b}\right)$ for polymorphs *a* and *b* phases (for the full derivation, see Supplementary Information, section 1). This concept also reduces reliance on the kinetic pre-factor $A$, which is primarily determined by the collision rate based on ideal gas kinetics. By focusing on the relative rates, the complex statistical treatments typically required to assess this term are canceled out.[27] Consequently, this framework enables efficient and straightforward predictions without the need for computationally intensive simulations of the pre-exponential terms.

To implement our hypothesis, we developed theoretical workflows for nucleation calculations (for details, see Supplementary Information, Section 2 and Figure S1). The key input parameters are energies of the bulk, surface, interface, epitaxial strain, and reaction energies. We determined these values using density functional theory and reaction networks[28], supplemented by literature data. Further details are provided in the Methods section.

***Precursor chemistry and polymorph selection in vapor deposition***

The $Ga_2O_3$ epitaxial system serves as a prototypical testbed for our framework owing to its rich polymorphism[29] and the availability of extensive experimental data. We first validated our model by addressing the long-standing issue of orientation inhomogeneity of β-$Ga_2O_3$ during homoepitaxial growth (Figure 2). Experiments show that even on the same β-$Ga_2O_3$ (001) substrate, the choice of precursor can dictate the film orientation: TMGa typically leads to (-201)-oriented β-phase, whereas TEGa favors the (001) orientation.[30]

The lattice mismatch between the (-201) and the (001) orientations is approximately 18% (Figure S2), which typically suppresses nucleation of (-201)-oriented domains. Nevertheless, the (-201) surface exhibited a substantially lower surface energy than the (001) surface, by 61 meV/$Å^2$, consistent with previous reports.[31, 32] This surface-energy gain can offset the strain penalty during epitaxy, thereby reducing the nucleation barrier via surface stabilization. Such stabilization can also be modulated by reaction-pathway energetics, as dictated by precursor control. We therefore hypothesize that the highly reactive pathway enabled by the TMGa precursor overcomes the structural penalty to stabilize the non-native (-201) orientation (Figure 2a).

By explicitly accounting for these energetic contributions, including surface stability, strain penalty, and reaction energies, our nucleation framework successfully predicted the experimental observations (See details in Supplementary Information, Section 3). In Figure 2b, we confirmed that the competitive nucleation between the two orientations undergoes a clear transition depending on $\Delta G_{rxn}$. Specifically, the TMGa-based pathway ($\Delta G_{rxn}$ = −359 meV/atom) crosses this transition point, enabling the (-201) orientation to outpace the (001) nucleation. In contrast, the TEGa-based pathway ($\Delta G_{rxn}$ = −195 meV/atom) remains confined to the (001) nucleation regime. These results directly identify precursor selection as a primary kinetic driver for orientation selection in homoepitaxy, providing a physical basis for controlling growth modes.

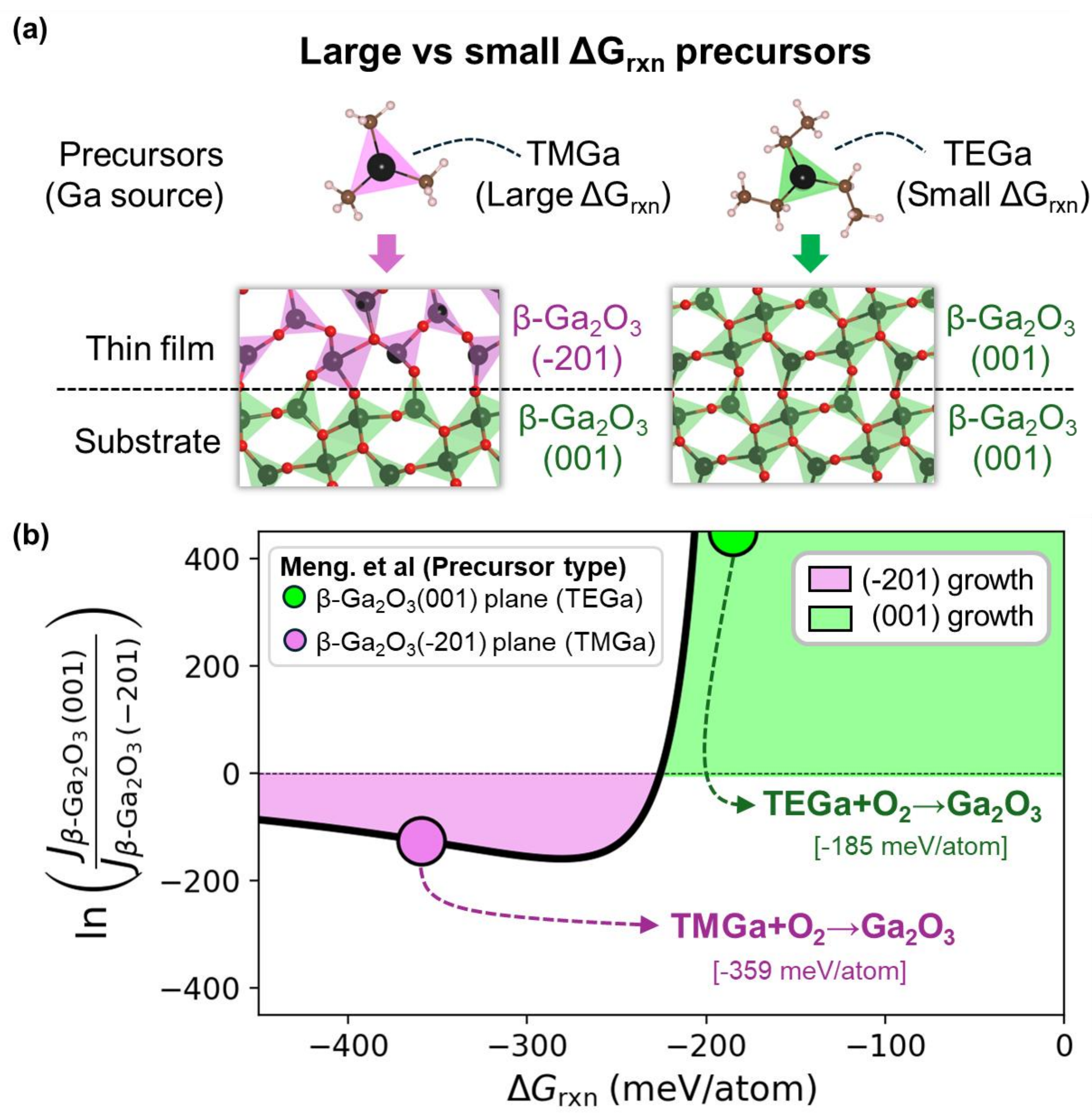

**Figure 2. Nucleation prediction on abnormal orientation growth in homoepitaxial *β*-$Ga_2O_3$.** (a) Summary of orientation competition illustrating kinetic regimes defined by precursor selection (TMGa vs TEGa), reconstructed from literature reports.[30] Calculated relative nucleation rate as a function of $\Delta G_{rxn}$ for (b) the (-201) vs (001) orientations with the precursor reaction points.

Extending this framework to heteroepitaxy on *α*-$Al_2O_3$ reveals a complex competitive landscape involving the *α*, *β*, and *κ* phases (Figure 3). On this substrate, *α*-$Ga_2O_3$ formation is unavoidable due to epitaxial stabilization.[33] The key competition, however, emerges only after this initial phase is formed. Once the *α*-$Ga_2O_3$ layer acts as a buffer layer, the lattice mismatch for the $\beta$ and $\kappa$ phases is substantially reduced compared to the bare substrate (See SI S4). This reduction attenuates the dominance of epitaxial strain, rendering phase selection increasingly sensitive to subtler contributions from interface interactions[34], surface stability[32], and vibrational energetics[35], in addition to residual strain effects (See SI S5 and S6).[17] Encouragingly, experiments show that this multi-factor competition remains amenable to precursor control, and we summarize these observations into the schematic shown in Figure 3a.

To rationalize the experimentally observed precursor dependence[33, 36, 37], we first examined α–β competition by predicting $\ln\left(\frac{J_{\beta-Ga_2O_3}}{J_{\alpha-Ga_2O_3}}\right)$ under the reported experimental conditions (Supplementary Information, Section 7). We identified that a clear transition in the preferred phase occurs only on the $\alpha$ buffer layer and is governed by $\Delta G_{rxn}$ (Figure S7). Mapping the experimentally reported precursor conditions onto this buffer-layer landscape, our model assigned the GaCl route to α-$Ga_2O_3$ and the TMGa route to β-$Ga_2O_3$, consistent with experiment (Figure 3b). The GaCl-based pathway ($\Delta G_{rxn}$ = −1040 meV/atom) yielded a markedly higher nucleation rate for the $\alpha$ phase, whereas the TMGa-based pathway ($\Delta G_{rxn}$ = −57 meV/atom) shifted the balance toward $\beta$. This mirrors the homoepitaxy case, where phase selection is driven by precursor selection.

By contrast, in the κ–β regime, phase selection is not dictated by the precursor selection but by process knobs that tune the precursor supersaturation.[38] In practice, the supersaturated condition is controlled by regulating the chamber pressure, which determines the partial pressure (P) of precursors. We calculated $\ln\left(\frac{J_{\beta-Ga_2O_3}}{J_{\kappa-Ga_2O_3}}\right)$ as a function of $\Delta G_{rxn}$ and delineated the phase boundary, guided by the experimentally reported data points (Supplementary Information, Section 8). Figure 3c shows that our nucleation predictions are consistent with $\kappa$-$Ga_2O_3$ formation within an intermediate chamber pressure range (20–50 mbar, corresponding to $\Delta G_{rxn}$ = −42 meV to −47 meV/atom).

However, at very high supersaturation (400 mbar, $\Delta G_{rxn}$ = −57 meV/atom), the thermodynamically stable $\beta$ phase was observed instead of $\kappa$, indicating a deviation from our prediction. We attribute this deviation to significant gas-phase pre-reactions and subsequent homogeneous nucleation occurring before the precursors reach the substrate, which circumvents epitaxial stabilization. Furthermore, at very low supersaturation (3 mbar, $\Delta G_{rxn}$ = −33 meV/atom), the reduced driving force is insufficient to sustain crystallization, leading to amorphization. To avoid these pitfalls, we constructed a T–P diagram that maps the predictable pressure (SI, Section S9).

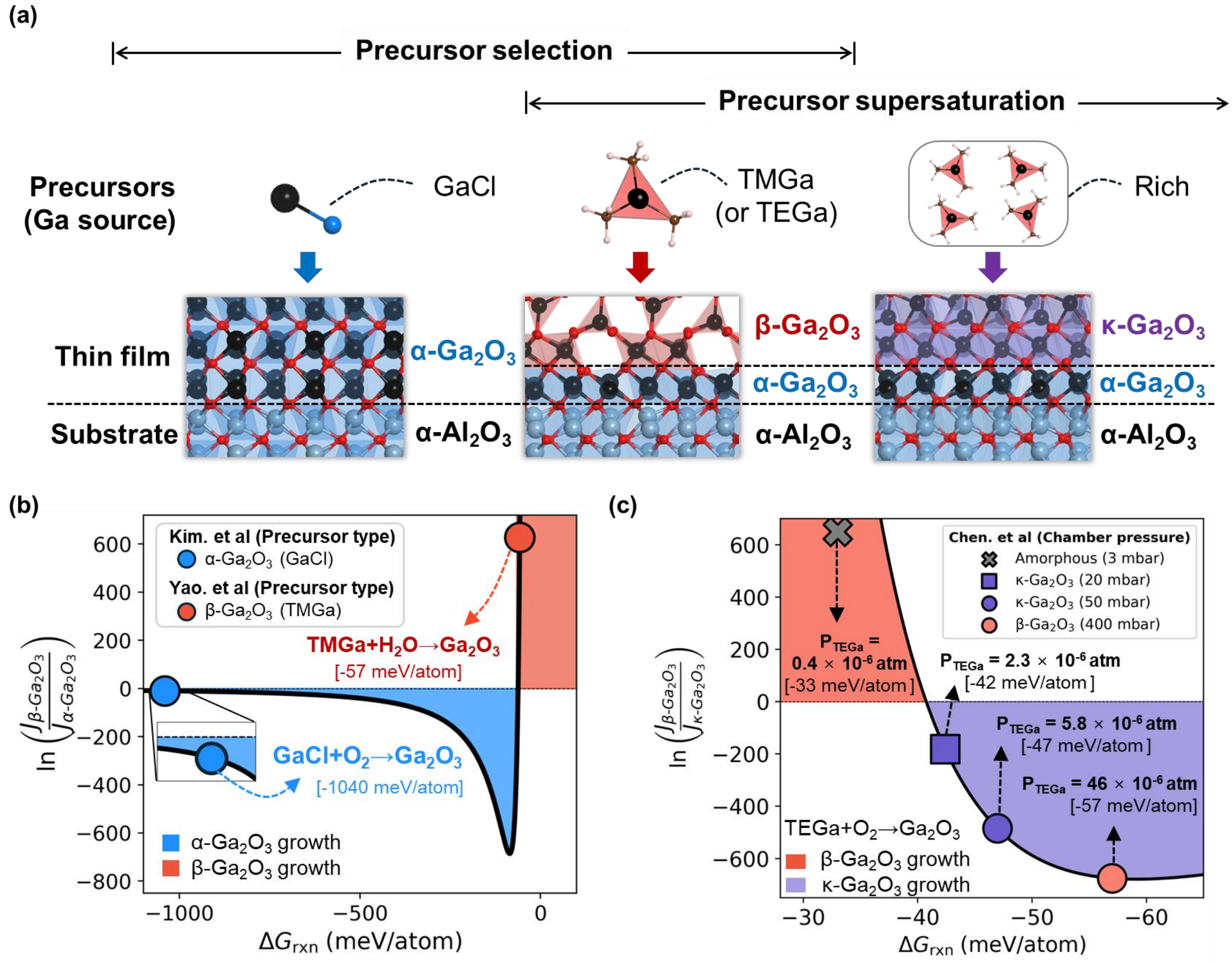

**Figure 3. Nucleation prediction on heteroepitaxial $Ga_2O_3$ on α-$Al_2O_3$ substrate.** (a) Summary of polymorphic competition illustrating two distinct kinetic regimes defined by precursor-controlled selection ($\alpha$ vs $\beta$) and flow-rate-controlled selection ($\beta$ vs $\kappa$), reconstructed from experimental reports.[33, 36-38] Calculated relative nucleation rate as a function of $\Delta G_{rxn}$ for (b) the $\alpha$ vs $\beta$ and (c) $\beta$ vs $\kappa$ with the precursor points.

### *Nucleation-based phase diagram of $Ga_2O_3$ deposition*

We constructed a nucleation phase diagram by combining experimental observations[37, 39] with our framework, as shown in Figure 4a (Supplementary Information, Section 10). These results collectively show that phase selection is governed not simply by relative thermodynamic stability, but by a coupled kinetic design in which the precursor-controlled reaction driving force. At the low-temperature regime (< about 450℃), crystallization is kinetically inhibited, leading to amorphous deposits. At high temperatures (above ~800 ℃), the thermodynamic ground state ($\beta$ phase) dominates regardless of precursor control. Crucially, the metastable $\alpha$- and $\kappa$-phases emerge exclusively within the intermediate temperature window of 450–650℃. This temperature dependence underscores that phase selection is not merely a function of $\Delta G_{rxn}$, but necessitates a precise kinetic design to ensure the reaction occurs within the surface-mediated nucleation zone.

Figure 4b elucidates the physical origin of this "polymorph synthesis window." At high temperatures, the excessive thermal energy triggers premature decomposition of precursors in the gas phase (pre-reaction) leading to homogeneous nucleation. In this freestanding limit where substrate effects are absent, bulk thermodynamics governs the phase selection, inevitably driving the system toward the stable β-phase. On the other end of the spectrum, insufficient driving force or thermal energy immobilizes adatoms, preventing them from reaching lattice sites and thereby yielding amorphous films. Therefore, realizing metastable polymorphs requires strictly confining the process to the "competition region" where surface-mediated heterogeneous nucleation prevails. In this kinetic window, the reaction remains distinct from the gas phase while retaining sufficient mobility to avoid amorphization, allowing the substrate- and precursor-driven stabilization of the desired metastable phases.

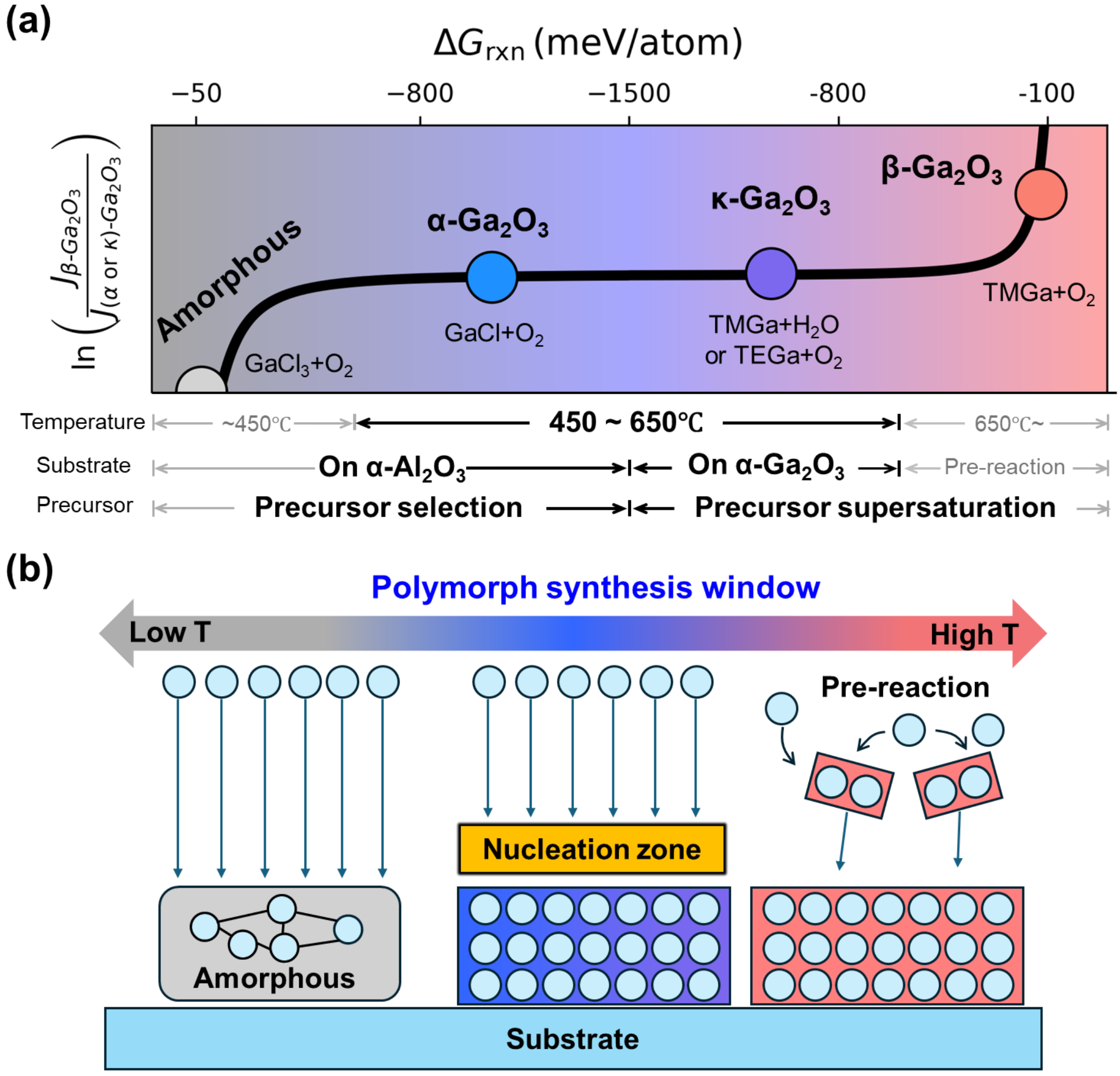

**Figure 4. Nucleation-based phase diagram and synthetic window for $Ga_2O_3$ polymorph.** (a) Nucleation phase diagram constructed by combining experimentally observed growth outcomes with our nucleation-based framework. Temperature, substrate, and precursor regimes are also delineated using the same procedure described above. (b) Schematic illustrating the physical origin of the synthetic window of polymorphs. At high temperature, precursors undergo gas phase pre reaction, promoting homogeneous nucleation. At low temperature, insufficient mobility prevents adatoms from reaching lattice sites, leading to amorphization. Metastable polymorphs are therefore accessible only in the intermediate competition region.

***Generalization to the $TiO_2$ system with rutile and anatase phases***

To demonstrate the universality of precursor-controlled nucleation model, we extend our analysis to $TiO_2$. Experiments show a clear precursor dependence in polymorph selection: TDMAT preferentially yields the metastable anatase phase, whereas TTIP favors the ground-state rutile phase.[40, 41] Applying our nucleation model, we computed the relative nucleation rates using values obtained from density functional theory calculations and literature values and confirmed that the predicted nucleation balance exhibits a clear transition with $\Delta G_{rxn}$ (Supplementary Information, Section 11). With experimental data, we calculated the precursor reaction energetics and used them to predict the nucleation behavior (Figure 5). TDMAT exhibits a negative $\Delta G_{rxn}$ (−240 meV/atom), shifting the predicted nucleation-rate balance toward anatase, whereas TTIP shows a less $\Delta G_{rxn}$ (−140 meV/atom) and favors the stable rutile phase. Together, this validation supports that our framework can be generalized across material systems. The TTIP precursor exhibits a mild reaction energy of $\Delta G_{rxn}$ -140 meV/atom, which provides insufficient driving force to overcome the barrier for the metastable state leading to the formation of the stable rutile phase. In contrast, the highly reactive TDMAT precursor with $\Delta G_{rxn}$ -240 meV/atom generates a substantial driving force that kinetically lowers the barrier for the metastable anatase structure. This successful prediction confirms that $\Delta G_{rxn}$ serves as a kinetic handle for phase selection across different material systems further validating the predictive power of our nucleation-based framework.

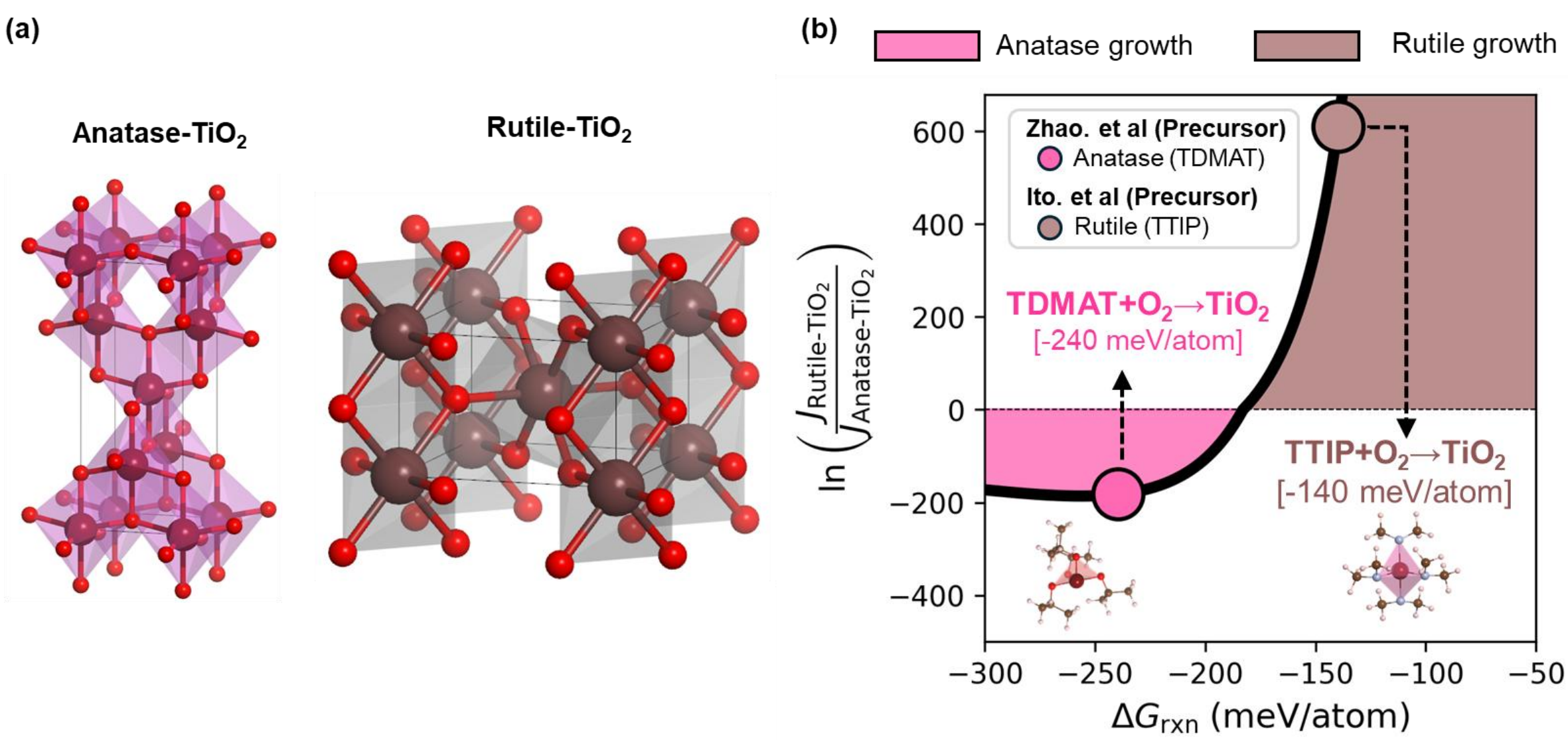

**Figure 5. Generalization of the nucleation framework to $TiO_2$ heteroepitaxy.** (a) Atomic structure of $TiO_2$ polymorphs. (b) Calculated relative nucleation rates of anatase- and rutile-$TiO_2$. The data points corresponding to the TDMAT and TTIP precursors were taken from experimental reports[40, 41].

## Conclusion

Metastable polymorph synthesis in vapor deposition has historically been largely serendipitous due to the complex interplay between thermodynamic stability and nucleation kinetics. In this work, we bring this empirical discovery process into a predictive scientific regime by establishing a nucleation-based framework that quantitatively links precursor reaction energetics, substrate interactions, and synthesis conditions to polymorph accessibility. By integrating first-principles energetics with CNT, we identify the reaction driving force as a key descriptor governing nucleation kinetics and phase competition. Applications to the $Ga_2O_3$ system successfully capture precursor-dependent orientation selection and polymorphic competition among the α, β, and κ phases, while extension to the $TiO_2$ system further confirms the generality of the approach. Collectively, we demonstrate that precursor chemistry provides a powerful chemical lever for controlling nucleation kinetics and polymorph selection during vapor deposition. These insights establish a predictive design principle for metastable polymorph synthesis and provide a pathway toward synthesis-by-design in complex materials systems. By transforming metastable-phase synthesis from empirical trial-and-error into predictive design, this work systematically unlocks hidden polymorph space across a broad chemical space and diverse synthesis platforms.

**Acknowledgements**

This work was supported by the Global-Learning & Academic Research Institution for Master's, PhD students, Postdocs (G-LAMP) program of the National Research Foundation of Korea (NRF) grant funded by the Ministry of Education (No. RS-2023-00285390). This work is also supported by the Ministry of Science and ICT (No. RS-2024-00444182). Computational resources were supported by the Korea Supercomputing Center (No. KSC-2023-CRE-0387, KSC-2024-CRE-0467 and No. KSC-2024-CRE-0049).

# Supplementary Information

# Precursor-Dependent Reaction Energetics as a Predictive Principle for Polymorph Selection in Thin Films

Hyeon Woo Kim[1,2], Han Uk Lee[1,3], Rohan Mishra[2*] & Sung Beom Cho[1,3,4*]

[1]Department of Material Science and Engineering, Ajou University, Suwon, Gyeonggi-do 16499, Republic of Korea

[2]Department of Mechanical Engineering & Material Science, Washington University in St. Louis, St. Louis, Missouri 63130, United States

[3]Department of Energy Systems Research, Ajou University, Suwon, Gyeonggi-do 16499, Republic of Korea

[4]School of Advanced Materials Science and Engineering, Sungkyunkwan University, Suwon, Gyeonggi-do 16419, Republic of Korea

*Corresponding authors.

Email: R. Mishra: rmishra@wustl.edu , S.B. Cho: csb@ajou.ac.kr

**Supplementary Method**

**First-principles calculations**

All calculations were performed using Vienna Ab initio Simulation Package (VASP) with the projector augmented-wave method with the generalized gradient approximation (GGA) within the Perdew–Burke–Ernzerhof (PBE) framework [1-3]. The plane-wave basis set was expanded to a cutoff energy of 520 eV. The structural optimization was truncated until the Hellmann–Feynman forces were under 0.01 eV Å$^{-1}$. The electronic energy convergence was set to 1 × 10$^{-5}$ eV. The Brillouin zone was sampled using an automatically generated k-point mesh with a reciprocal density of 100/Å$^3$ using *pymatgen* [4]. For the slab models, a vacuum spacing of at least 15 Å of was inserted to minimize interactions between periodic images. Gas molecules were calculated within a 15 Å ×15 Å ×15 Å cubic box to prevent spurious intermolecular interactions. For $TiO_2$ calculations, the GGA+U approach was employed with an effective Hubbard parameter of U = 7 eV applied to the Ti 3d states [5].

**Reaction list and energetics**

Reaction modeling was performed using a graph-based Reaction Network [6]combined with first-principles-calculated energies. In vapor deposition, each reaction pathway was formulated by combining precursors, reactants, products, and by-products. Each chemical species was defined as an entry with its computed total energy, and the entries were categorized according to the metal-source precursors. All possible reactions were enumerated using *BasicEnumerator* algorithm, reactions with unbalanced stoichiometry were excluded.

The reaction energy is defined as:

$$\Delta G_{rxn}^{0} = \sum_{i \in products} \nu_i \mu_i^0 - \sum_{j \in precursors} \nu_j \mu_j^0$$

where $\nu_i$ and $\nu_j$ are the stoichiometric coefficients, $\mu_i^0$ and $\mu_j^0$ are the first-principles total energies of products and gas phase precursors, respectively. To implement the realistic process condition, the ideal gas approximation was applied to evaluate the temperature- and pressure-dependent chemical potentials of each gaseous species as follows:

$$\Delta\mu_j(T, p_j) = u_j^0 + k_B T ln\left(\frac{p_j}{p_j^0}\right)$$

where $k_B$ is the Boltzmann constant, $T$ is temperature, $p_j$ is the partial pressure of gas-phase precursors, and $p_j^0$ is the reference pressure. These temperature- and pressure-calculated chemical potentials were included in the reaction energy calculations to account for realistic deposition process conditions [7].

**Helmholtz free energy calculation**

The $E_{helm}$ was derived from finite-temperature Gibbs free energy using *Phonopy* package [8]. The effect of thermal expansion was incorporated through the quasi-harmonic approximation (QHA), which combines phonon and total energy calculations at various fixed volumes $(V)$. The Gibbs free energy at a given temperature and constant pressure $(p)$ was evaluated by minimizing the free energy with respect to volume, expressed as:

$$G(T,p) = min_V[E(V,0K) + F_{phonon} + pV]$$

where $E(V,0K)$ is total energy without vibrational contributions and $F_{phonon}$ is the phonon free energy, resulting in the equilibrium volume at each temperature. All supercells were constructed with lattice constants of 10 Å or larger (or similar). To apply QHA, the cell volumes were expanded and compressed with linear strains ranging from -5% to +5% in 1% increments.

**Epitaxial strained energy calculation**

The $E_{epi}$ was computed using a combination of coincident-site lattice models (CSL) and first-principles calculations [9]. According to CSL theory, the epitaxial interface is defined such that a periodic repeating unit is formed where the lattice sites of the film and substrate coincide. Although multiple CSL configurations can exist for a given film–substrate pair, the smallest CSL is generally favored as it reduces the number of dangling bonds. For $Ga_2O_3$ on $\alpha$-$Al_2O_3$ substrate, the preferred lattice orientations were used as (-201) for β phase, (001) for α phase, and (001) for ε phase, based on comprehensive experimental studies [10]. For $TiO_2$ on $SrTiO_3$ substrate, previously reported epitaxial strain energies were used collected from the literature due to their well-established consistency with experimental and theoretical data [11].

**Surface energy calculation**

The $\gamma_{surf}$ was determined for fully relaxed slab models using the following relation:

$$\gamma_{surf} = \frac{1}{2A}(E_{slab} - \frac{N_{slab}}{N_{bulk}}E_{bulk})$$

where $A$ is the surface area, and $E_{slab}$ and $E_{bulk}$ are the total energies of the slab supercell and the bulk primitive cell for each polymorph, respectively. $N_{slab}$ and $N_{bulk}$ are the number of atoms in the slab and bulk cells. For $Ga_2O_3$ polymorphs, the (-201), (001), and (001) planes, aligned with preferred lattice orientation in epitaxy, were considered for the β, α, and ε phases, respectively. Because the $\beta$- and $\kappa$-$Ga_2O_3$ surfaces exhibit higher atomic anisotropy and polarity compared to the $\alpha$-phase, surface reconstruction was considered, and the corresponding energetic contributions were referenced from previously reported literature data [12]. For $TiO_2$ polymorphs, the (111) rutile and (001) anatase planes, consistent with their preferred epitaxial orientations, were adopted, and the surface energies were obtained from literature-reported values [11].

**Interface energy**

The $\gamma_{inter}$ was obtained by accounting for the epitaxial relationship between the substrate and the preferred orientations of the polymorphs, with reference to previously reported literature values [11, 13]. Direct first-principles calculation of interface energetics is computationally demanding because it requires constructing large supercells that simultaneously capture atomic relaxations, lattice mismatches, and interfacial dipoles between two distinct phases [14]. Moreover, these calculations are often prohibitively expensive due to convergence with respect to slab thickness, vacuum spacing, and diversity of atomic terminations [15]. Considering these challenges, as well as the availability of extensive benchmark data for similar oxide interfaces [16], the use of literature-reported values provides a reliable approximation for the present analysis.

## S1. Detailed description of relative nucleation rates in vapor deposition

The relative nucleation rate is defined as the ratio of the nucleation rates of two competing polymorphs, $x$ (stable) and $y$ (metastable), which may form when a given material $AO$ grows on a substrate $Z$. As described in the main text, the reformulated nucleation rate of $AO$ on substrate $Z$ is defined as follows:

$$J_{AO}^{Z} = A_{AO} exp\left[-\frac{16\pi(\gamma_{surf,AO} + \gamma_{inter,AO}^{Z})^3}{3n_{AO}^2 k_B T(\Delta G_{rxn,AO} + E_{epi,AO}^{Z})^2}\right] \tag{S1.1}$$

where $A_{AO}$ is a pre-factor of $AO$, $n_{AO}$ is atomic density of $AO$, $k_B$ is Boltzmann's constant, $T$ is temperature, $\gamma_{surf,AO}$ is surface energy of $AO$, $\gamma_{inter,AO}^{Z}$ is interface energy between $AO$ and $Z$, $E_{epi,AO}^{Z}$ is epitaxial strain energy of $AO$ on $Z$ substrate, and $\Delta G_{rxn}, AO$ is the reaction energy of $AO$ from the precursor states. Here, $\gamma_{surf,AO} + \gamma_{inter,AO}^{Z}$ and $n_{AO}\left(\Delta G_{rxn,AO} + E_{epi,AO}^{Z}\right)$ represent the $\gamma_{surf}$ and $\Delta G_{bulk}$ terms of the original CNT equation, respectively. Among these factors, $n_{AO}$, $\gamma_{surf,AO}$, $\gamma_{inter,AO}^{Z}$, and $E_{epi,AO}^{Z}$ are polymorph dependent. For stable polymorph $x$, $J_{x-AO}^{Z}$ is expressed as follows:

$$J_{x-AO}^{Z} = A_{x-AO} exp\left[-\frac{16\pi(\gamma_{surf,x-AO} + \gamma_{inter,x-AO}^{Z})^3}{3(n_{x-AO})^2 k_B T(\Delta G_{rxn,AO} + E_{epi,x-AO}^{Z})^2}\right] \tag{S1.2}$$

In the relative nucleation rate, $\Delta G_{rxn,AO}$ contributes bulk energy but does not distinguish between polymorphs, even though bulk energy differences must be considered. To this end, we employed relative energy differences, defined as $\Delta E = E_y - E_x$ for the Helmholtz free-energy including vibrational entropy, ($\Delta E_{helm}$) and the epitaxially strained formation energies ($\Delta E_{epi}^{Z}$), respectively. Therefore, the relative nucleation rate for $x$ and $y$ is expressed as follows:

$$\frac{J_{x-AO}^{Z}}{J_{y-AO}^{Z}} = \frac{A_{x-AO}}{A_{y-AO}}\left[exp\frac{16\pi}{k_B T}\left(\frac{(\gamma_{surf,y-AO} + \gamma_{inter,y-AO}^{Z})^3}{(n_{y-AO})^2(\Delta G_{rxn,AO} + \Delta E_{helm} + \Delta E_{epi}^{Z})^2} - \frac{(\gamma_{surf,x-AO} + \gamma_{inter,x-AO}^{Z})^3}{(n_{x-AO})^2(\Delta G_{rxn,AO})^2}\right)\right] \tag{S1.3}$$

In gas-to-solid transformations, two distinct crystallization routes are possible: an amorphous-to-crystal pathway characterized by slow kinetics (e.g., atomic layer deposition), and a direct vapor-to-crystal pathway governed by fast gas-phase kinetics (e.g., metal–organic chemical vapor deposition and halide vapor phase epitaxy). In the former case, kinetic factors (e.g., attachment and detachment) play a dominant role in determining crystallization behavior. In contrast, the latter can often be approximated as ideal-gas-like system, given that the partial pressures in the chamber are typically sufficiently high compared to those in amorphous-mediated growth. In this work, we consider the latter regime, as the vapor deposition conditions of interest involve sufficiently high precursor fluxes and chamber partial pressures such that gas-phase supply is not rate-limiting for nucleation. Under these conditions, the ratio of the pre-factors $\left(\frac{A_{x-AO}}{A_{y-AO}}\right)$ can be neglected, allowing the relative nucleation rate to be expressed as follows:

$$\frac{J_{x-AO}^{Z}}{J_{y-AO}^{Z}} = exp\frac{16\pi}{k_B T}\left(\frac{(\gamma_{surf,y-AO} + \gamma_{inter,y-AO}^{Z})^3}{(n_{y-AO})^2(\Delta G_{rxn,AO} + \Delta E_{helm} + \Delta E_{epi}^{Z})^2} - \frac{(\gamma_{surf,x-AO} + \gamma_{inter,x-AO}^{Z})^3}{(n_{x-AO})^2(\Delta G_{rxn,AO})^2}\right) \quad (S1.4)$$

Owing to the exponential dependence of the nucleation rate, comparisons are most meaningfully made on a logarithmic scale, yielding the following expression:

$$ln\left(\frac{J_{x-AO}^{Z}}{J_{y-AO}^{Z}}\right) = \frac{16\pi}{k_B T}\left(\frac{(\gamma_{surf,y-AO} + \gamma_{inter,y-AO}^{Z})^3}{(n_{y-AO})^2(\Delta G_{rxn,AO} + \Delta E_{helm} + \Delta E_{epi}^{Z})^2} - \frac{(\gamma_{surf,x-AO} + \gamma_{inter,x-AO}^{Z})^3}{(n_{x-AO})^2(\Delta G_{rxn,AO})^2}\right) \quad (S1.5)$$

## S2. Theoretical workflow to calculate relative nucleation rates

The input parameters used to calculate relative nucleation rates were classified into two categories, as shown in Figure S1. One category corresponds to polymorphic energetics, encompassing both intrinsic and extrinsic properties, while the other corresponds to reaction energetics, which serve as kinetic handles.

For intrinsic polymorphic energetics, finite-temperature free energies and surface energetics were determined in the absence of substrate effects. In contrast, extrinsic polymorphic energetics, including interface energies and strain energetics, were evaluated by explicitly accounting for substrate-induced effects. By incorporating these parameters, we calculate relative nucleation rates as a function of reaction energetics and evaluate nucleation competition by examining whether the relative rate is positive or negative.

To investigate kinetic handles by precursor control, reaction energetics were calculated. A reaction is defined as a set consisting of precursors, reactants, target products, and by-products. In vapor deposition processes, the concurrent formation of volatile by-products is unavoidable, resulting in a combinatorially large reaction-pathway space and rendering systematic pathway identification impractical. To address this challenge, we employed a graph-based reaction-network approach [6], which systematically generates all plausible reactions and evaluates their energetics by specifying reaction pools consisting of the initial reactants, target products, and a constrained yet process-relevant set of by-product species. This strategy enables the automated construction of reaction networks while systematically identifying process-relevant reaction pathways.

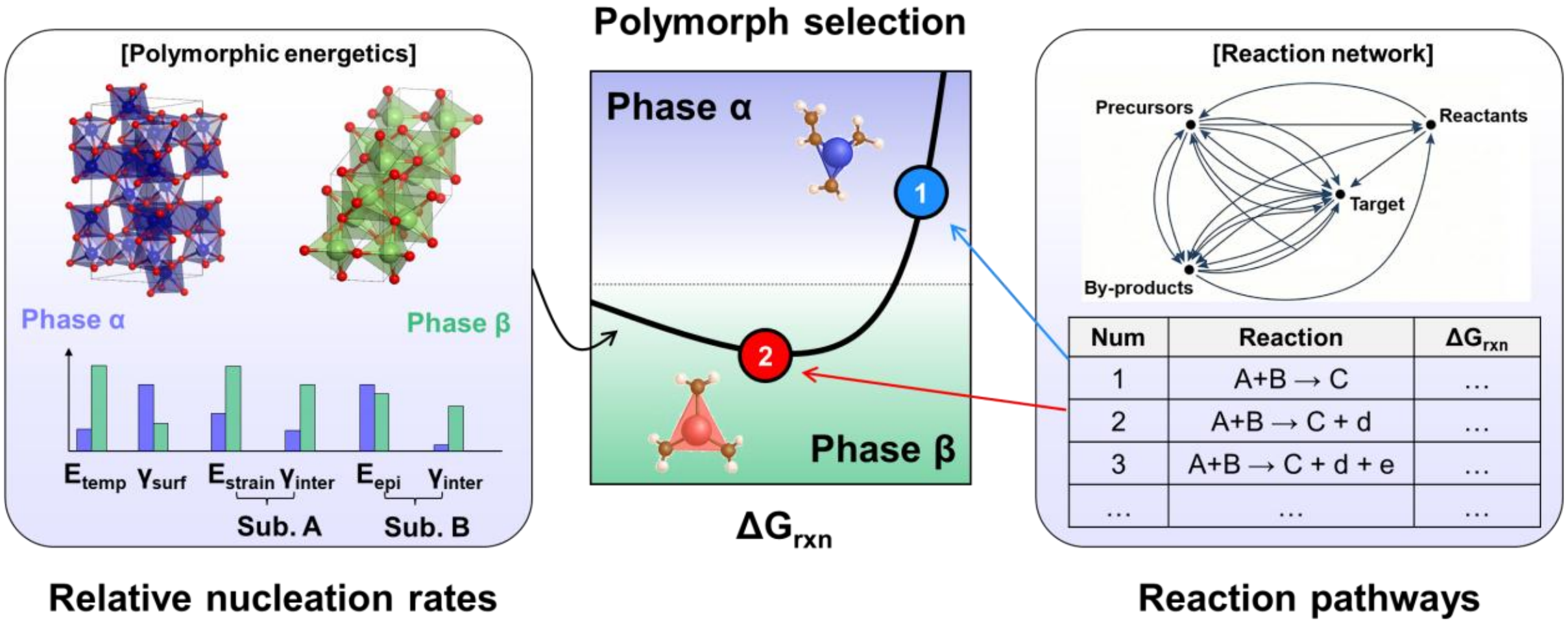

**Figure S1.** Workflow of the nucleation framework used to evaluate polymorph selection under vapor-phase deposition.

## S3. Prediction of homoepitaxial β-$Ga_2O_3$ nucleation

Experiments report that even on the same β-$Ga_2O_3$ (001) substrate, the choice of precursor can dictate the film orientation: TMGa typically leads to (-201)-oriented β-phase, whereas TEGa favors the (001) orientation [17].

To predict anomalous growth in homoepitaxial β-$Ga_2O_3$, the relative nucleation rates on (001) and (-201) orientations were calculated using DFT calculations. We computed the surface energies as $\gamma_{\mathrm{surf},\beta-\mathrm{Ga_2O_3}\,(001)} = 101.28\ \mathrm{meV/Å^2}$ and $\gamma_{\mathrm{surf},\beta-\mathrm{Ga_2O_3}\,(-201)} = 40.31\ \mathrm{meV/Å^2}$. For (-201) orientations, a surface reconstruction energy of −8 meV/atom was included, as this surface is sensitive to dangling-bond–dependent terminations [12]. The $n_{\beta-\mathrm{Ga_2O_3}}$ was taken as 0.0909 atom/$Å^3$ and temperature was set to 900 °C. The $\Delta E^{\mathrm{epi},\mathrm{B-Ga_2O_3(001)}}$ between (-201) and (001) was 169 meV/atom, obtained by explicitly modeling the lattice mismatch to -1.61% along the [010] direction and 18.29% along the [10-1] direction, as shown in Figure S2. For the homoepitaxial case, the interface energy and Helmholtz free-energy contributions were not considered, as nucleation occurs within a single phase. These contributions are explicitly considered for heteroepitaxial systems discussed below. By inputting calculated parameters to S1.5, we calculated relative nucleation rates, as shown in Figure S3.

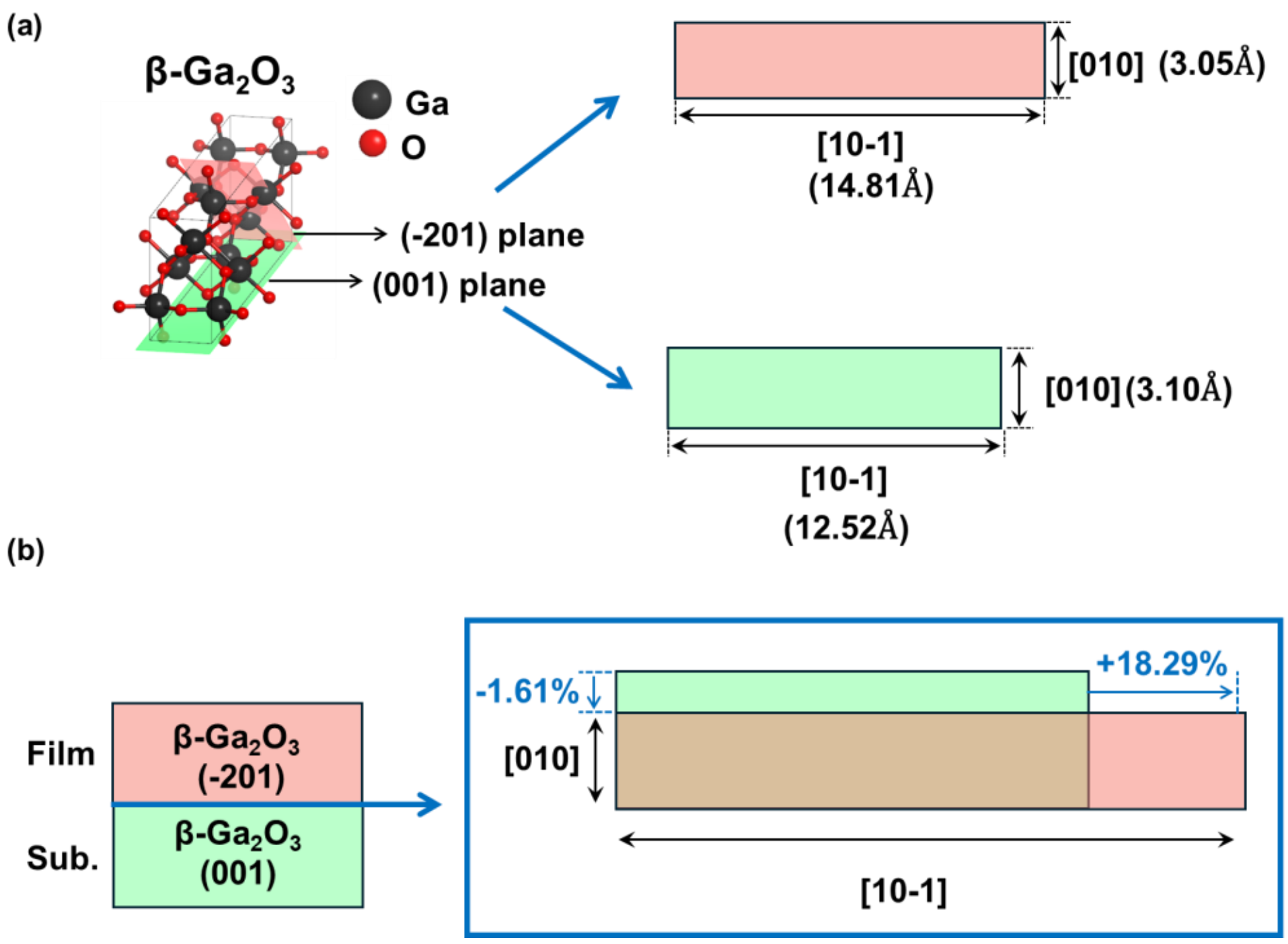

**Figure S2.** (a) (001) and (-201) surfaces of β-$Ga_2O_3$ and (b) the epitaxial strain of (-201) orientation on the (001) substrate.

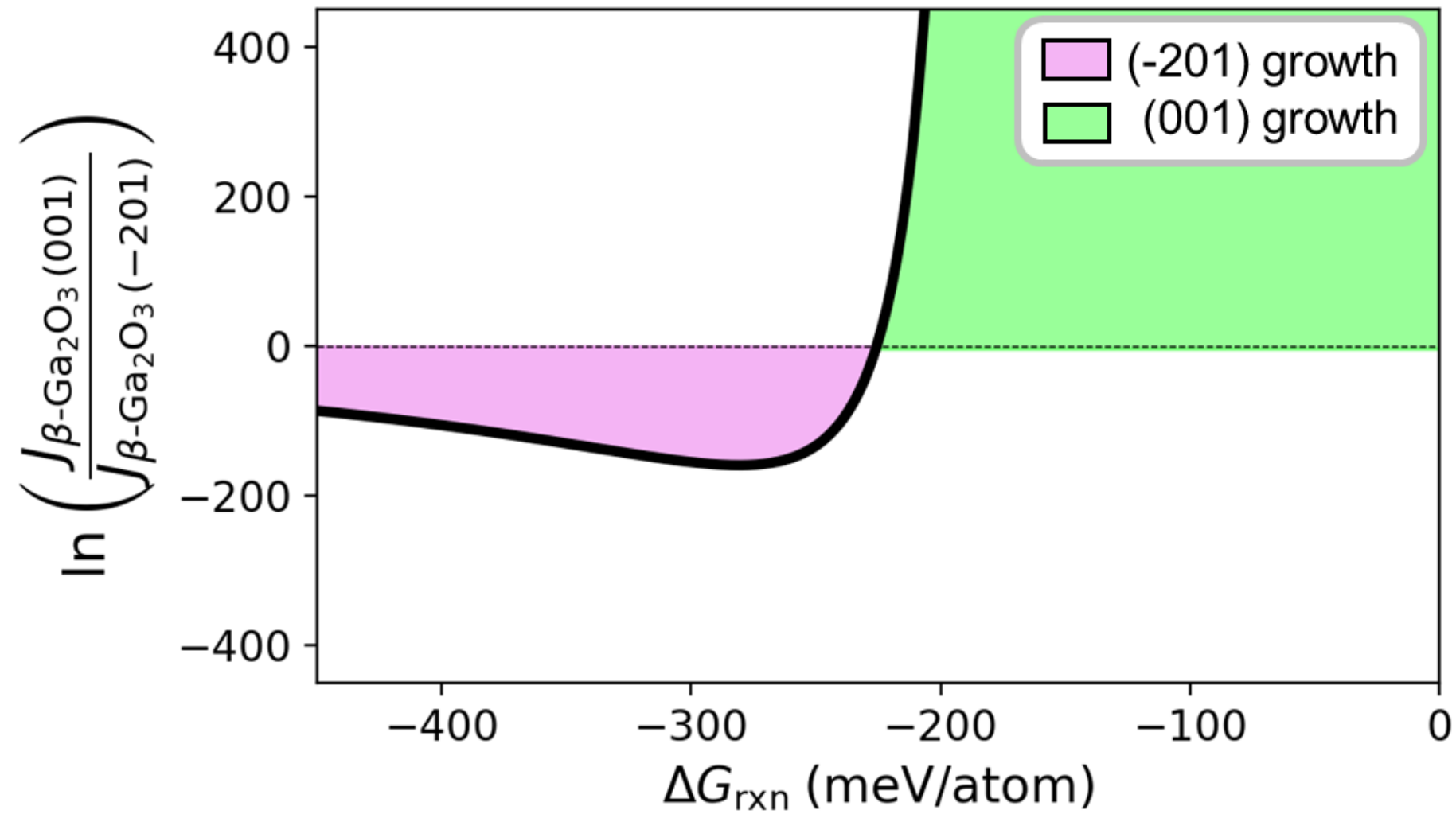

**Figure S3.** Calculated relative nucleation rates for the (001) and (-201) orientations in (001) β-$Ga_2O_3$ homoepitaxy

We next determine possible reactions in this process. We first investigated reaction pools from experimental literature. TMGa and TEGa were used for Ga precursors, $O_2$ used for reactions. Additionally, we considered intermediate source gas from used precursors and by products, and then as shown in Figure S4. Enumerated reaction list was summarized in Table S1. From experimental reports, we collected process conditions and calculated $\Delta G_{rxn}$ as follows:

※Condition: 900℃, $F_{O2}$ =800 SCCM, $F_{TEGa}$ = 31 μmol/min $F_{TMGa}$ = 58 μmol/min $P_{chamber}$ = 60 Torr
1. TEGa + $O_2$ → $Ga_2O_3$+$H_2O$+$CH_2$ ($\Delta G_{rxn}$ = -185 meV/atom)
2. TMGa + $O_2$ → $Ga_2O_3$+$H_2O$+$CH_2$ ($\Delta G_{rxn}$ = -359 meV/atom)

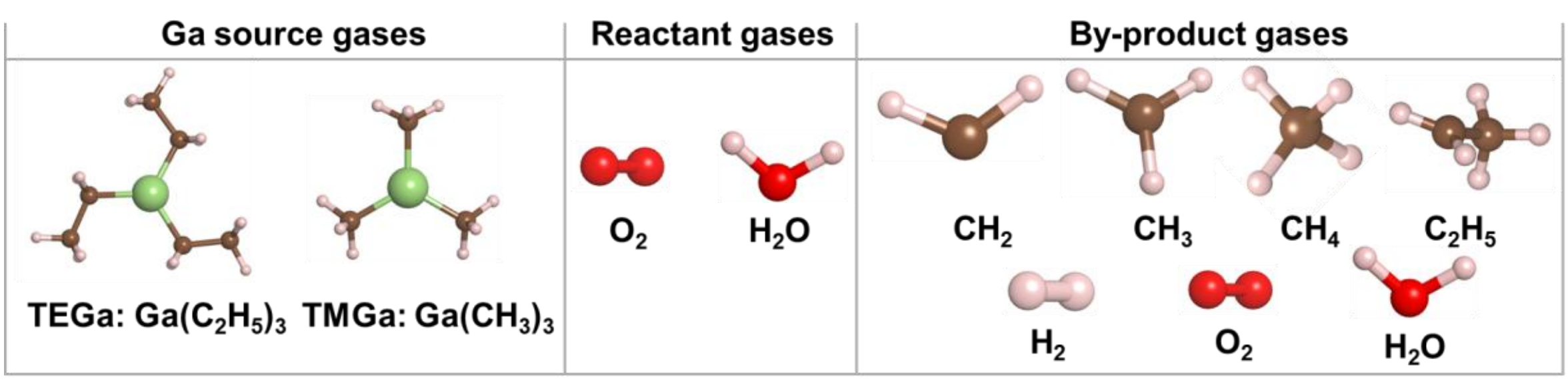

**Figure S4.** Possible reaction components in β-$Ga_2O_3$ homoepitaxy

| Reaction | Precursor | Reactants | By-products | ΔGrxn (meV/atom) | rxn |
|---|---|---|---|---|---|
| 1 | TMGa | O2 | H2C, H2O | -359.31 | 0.5 O2 + 0.3333 Ga(H3C)3 -> H2C + 0.1667 Ga2O3 + 0.5 H2O |
| 2 | TMGa | O2 | H5C2, H2O | -209.98 | 0.75 O2 + 0.6667 Ga(H3C)3 -> H5C2 + 0.3333 Ga2O3 + 0.5 H2O |
| 3 | TEGa | O2 | H2C, H2O | -185.44 | 0.1667 Ga(H5C2)3 + 0.25 O2 -> H2C + 0.08333 Ga2O3 + 0.25 H2O |
| 4 | TMGa | O2 | H2C, H2 | -151.36 | 0.25 O2 + 0.3333 Ga(H3C)3 -> H2C + 0.5 H2 + 0.1667 Ga2O3 |
| 5 | TMGa | O2 | H2, H5C2 | -98.28 | O2 + 1.333 Ga(H3C)3 -> H2 + 2 H5C2 + 0.6667 Ga2O3 |
| 6 | TEGa | O2 | H2C, H2 | -45.54 | 0.1667 Ga(H5C2)3 + 0.125 O2 -> H2C + 0.25 H2 + 0.08333 Ga2O3 |
| 7 | TEGa | O2 | H5C2 | 19.97 | 0.3333 Ga(H5C2)3 + 0.25 O2 -> H5C2 + 0.1667 Ga2O3 |
| 8 | TMGa | O2 | H3C | 100.62 | O2 + 1.333 Ga(H3C)3 -> 4 H3C + 0.6667 Ga2O3 |
| 9 | TEGa | O2 | H2C, H3C | 109.94 | 0.3333 Ga(H5C2)3 + 0.25 O2 -> H2C + H3C + 0.1667 Ga2O3 |
| 10 | TMGa | O2 | H5C2, H4C | 155.98 | 0.75 O2 + Ga(H3C)3 -> H5C2 + H4C + 0.5 Ga2O3 |
| 11 | TEGa | O2 | H2C, H4C | 206.16 | 0.2222 Ga(H5C2)3 + 0.1667 O2 -> H2C + 0.3333 H4C + 0.1111 Ga2O3 |
| 12 | TMGa | O2 | H2C, H4C | 256.57 | 0.5 O2 + 0.6667 Ga(H3C)3 -> H2C + H4C + 0.3333 Ga2O3 |
| 13 | TMGa | O2 | CO2, H2O | 429.23 | 2 O2 + 0.3333 Ga(H3C)3 -> CO2 + 0.1667 Ga2O3 + 1.5 H2O |
| 14 | TEGa | O2 | H3C, CO2 | 504.58 | 0.2 Ga(H5C2)3 + 0.35 O2 -> H3C + 0.2 CO2 + 0.1 Ga2O3 |
| 15 | TEGa | O2 | CO2, H2O | 655.46 | 0.1667 Ga(H5C2)3 + 1.75 O2 -> CO2 + 0.08333 Ga2O3 + 1.25 H2O |
| 16 | TMGa | O2 | H4C, CO2 | 785.93 | 0.6667 O2 + 0.4444 Ga(H3C)3 -> H4C + 0.3333 CO2 + 0.2222 Ga2O3 |
| 17 | TMGa | O2 | H2, CO2 | 1043.58 | 0.8333 O2 + 0.2222 Ga(H3C)3 -> H2 + 0.6667 CO2 + 0.1111 Ga2O3 |
| 18 | TEGa | O2 | H4C, CO2 | 1120.49 | 0.2667 Ga(H5C2)3 + 0.8 O2 -> H4C + 0.6 CO2 + 0.1333 Ga2O3 |
| 19 | TEGa | O2 | H2, CO2 | 1296.16 | 0.1333 Ga(H5C2)3 + 0.9 O2 -> H2 + 0.8 CO2 + 0.06667 Ga2O3 |

Table S1. Enumerated reaction list for $\beta$-$Ga_2O_3$ homoepitaxy.

## S4. Epitaxial strain of $Ga_2O_3$ polymorphs on α-$Al_2O_3$ substrate

The atomic structures and preferentially oriented planes of $Ga_2O_3$ on α-$Al_2O_3$ are shown in Figure S5 a-c. Using coincident-site lattice (CSL) theory [9], the possible epitaxial relationships are presented in Figure S5 d-e. In the growth of $Ga_2O_3$ on α-$Al_2O_3$, α-$Ga_2O_3$ has been reported to act as a buffer layer owing to its isostructural nature with the substrate [18]. Taking this effect into account, we evaluated the epitaxial strains, as summarized in Table S2.

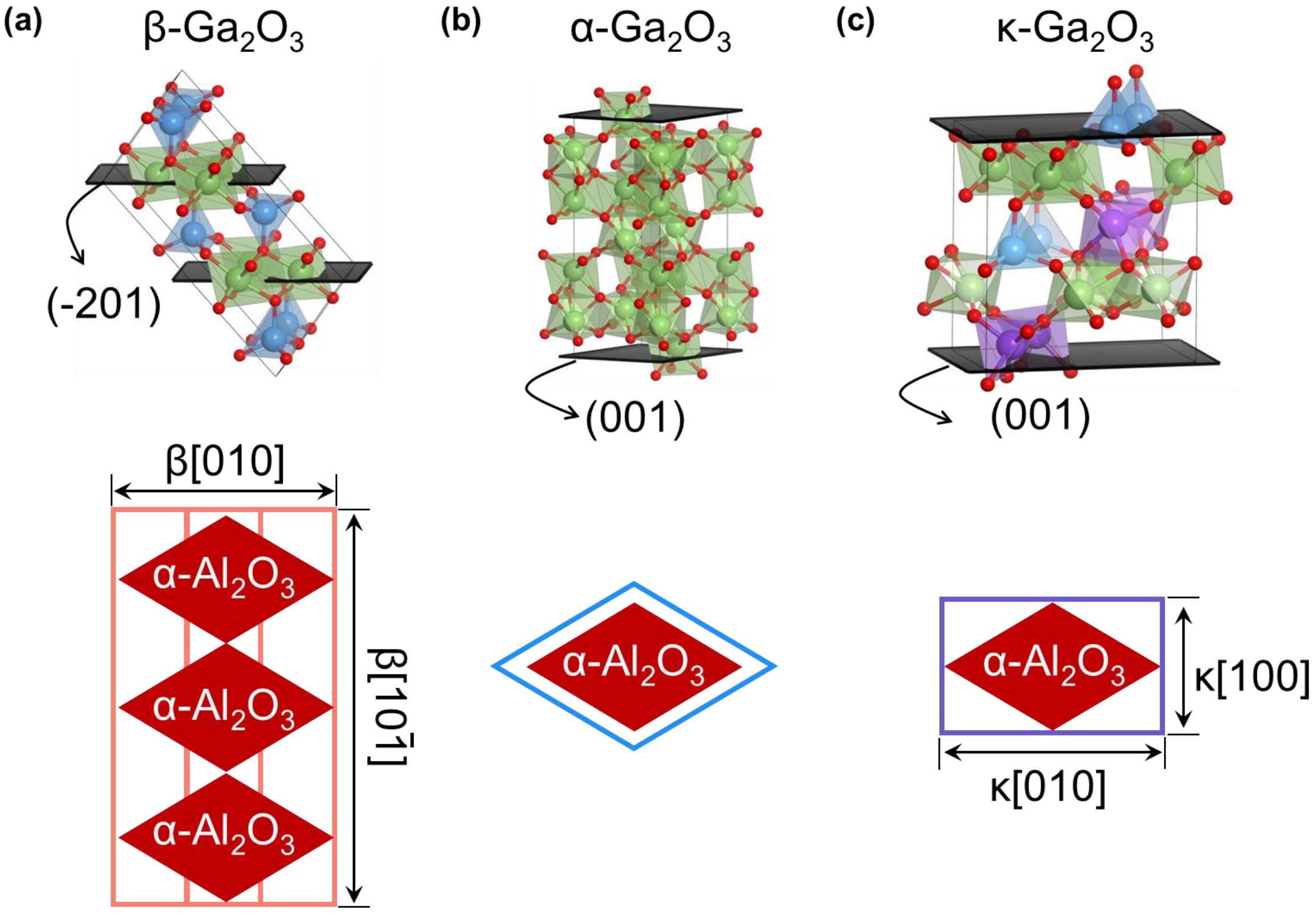

**Figure S5.** Atomic structure, the preferentially oriented planes, and CSL r of (a) α-, (b) β-, and (c) κ-$Ga_2O_3$ polymorphs on α-$Al_2O_3$ substrate. (d-f) show the possible CSL models.

| Material | α-$Al_2O_3$ | | α-$Ga_2O_3$ | |
|---|---|---|---|---|
| | strain,a | strain,b | strain,a | strain,b |
| β-$Ga_2O_3$ (-201) | -11.25% | -3.86% | -5.60% | 1.41% |
| α-$Ga_2O_3$ (001) | -5.35% | -5.35% | 0.00% | 0.00% |
| κ-$Ga_2O_3$ (001) | -6.64% | -5.76% | -1.22% | -0.39% |

**Table S2.** Calculated epitaxial strain in $Ga_2O_3$ polymorphs on α-$Al_2O_3$ substrate and α-$Ga_2O_3$ buffer layer.

**S5. Polymorphic energetics of α- and β-$Ga_2O_3$ polymorphs on α-$Al_2O_3$ substrate**

| Name | Value | Unit |
|---|---|---|
| $\gamma_{surf,\beta-Ga_2O_3\ (-201)}$ | 40* | meV/Å$^2$ |
| $\gamma_{surf,\alpha-Ga_2O_3\ (001)}$ | 62 | meV/Å$^2$ |
| $\gamma_{inter,\beta-Ga_2O_3\ (-201)}^{epi,\alpha-Al_2O_3(001)}$ | 53[a] | meV/Å$^2$ |
| $\gamma_{inter,\beta-Ga_2O_3\ (-201)}^{epi,\alpha-Ga_2O_3(001)}$ | 47[a] | meV/Å$^2$ |
| $\gamma_{inter,\alpha-Ga_2O_3\ (001)}^{epi,\alpha-Al_2O_3(001)}$ | -3[a] | meV/Å$^2$ |
| $n_{\beta-Ga_2O_3\ (-201)}$ | 0.091 | atom/Å$^3$ |
| $n_{\beta-Ga_2O_3\ (-201)}^{epi,\alpha-Al_2O_3(001)}$ | 0.100 | atom/Å$^3$ |
| $n_{\beta-Ga_2O_3\ (-201)}^{epi,\alpha-Ga_2O_3(001)}$ | 0.094 | atom/Å$^3$ |
| $n_{\alpha-Ga_2O_3\ (-201)}$ | 0.099 | atom/Å$^3$ |
| $n_{\alpha-Ga_2O_3\ (-201)}^{epi,\alpha-Al_2O_3(001)}$ | 0.106 | atom/Å$^3$ |
| $\Delta E_{epi}^{\alpha-Al_2O_3}$ | -41.5 | meV/atom |
| $\Delta E_{epi}^{\alpha-Ga_2O_3}$ | -0.7 | meV/atom |
| $\Delta E_{helm}$ | 26 | meV/atom |

**Table S3.** Determined parameters to calculate nucleation rates for α- and β-$Ga_2O_3$ competition. * indicates that the reconstruction contribution was included [12]. [a] [13]

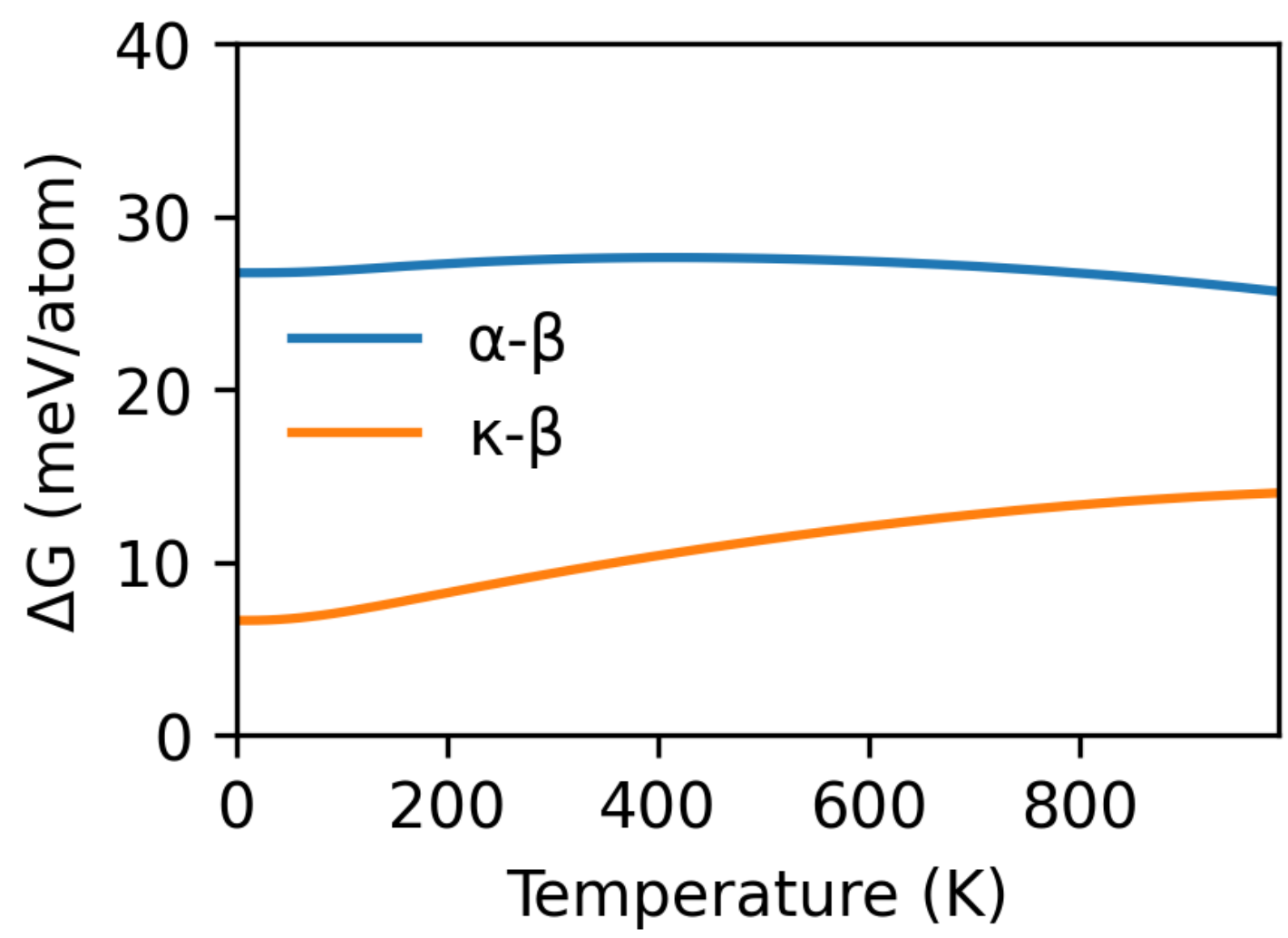

**Figure S6.** Calculated finite-temperature Gibbs free energy for α-, β-, and κ-$Ga_2O_3$ systems.

## S6. Polymorphic energetics of κ- and β-$Ga_2O_3$ polymorphs on α-$Ga_2O_3$ substrate

| Name | Value | Unit |
| --- | --- | --- |
| $\gamma_{surf,\kappa-Ga_2O_3\,(001)}$ | 45[1]* | meV/Å$^2$ |
| $\gamma^{epi,\alpha-Ga_2O_3(001)}_{inter,\kappa-Ga_2O_3\,(001)}$ | 32[2] | meV/Å$^2$ |
| $n_{\kappa-Ga_2O_3\,(001)}$ | 0.0939 | atom/Å$^3$ |
| $n^{epi,\alpha-Al_2O_3(001)}_{\kappa-Ga_2O_3\,(001)}$ | 0.0948 | atom/Å$^3$ |
| $\Delta E^{\alpha-Ga_2O_3}_{epi}$ | -5.57 | meV/atom |
| $\Delta E_{helm}$ | 13 | meV/atom |

**Table S4.** Determined parameters to calculate nucleation rates for κ- and β-$Ga_2O_3$ competition. * indicates that the reconstruction contribution was included [12].

## S7. Prediction of heteroepitaxial α- and β-$Ga_2O_3$ nucleation on α-$Al_2O_3$ substrate

By inputting Table S3 to S1.5, we calculated the relative nucleation rates shown in Figure S7. Based on experimental reports [10, 18, 19], we collected process conditions as follows:

HVPE conditions: 650℃, $F_{O2}$ =80 SCCM, $F_{GaCl}$ = 138 SCCM $P_{chamber}$ = 550 Torr

MOVPE conditions: 800℃, $F_{O2}$ =400 SCCM, $F_{TMGa}$ = 5 SCCM $P_{chamber}$ = 50 mbar

We then configured the set of plausible reaction components (Figure S8) and screened the following reactions:

1. HVPE: GaCl+ $O_2$ → $Ga_2O_3$ + $Cl_2$ ($\Delta G_{rxn}$ = -1040 meV/atom)
2. MOVPE: $Ga(CH_3)$ + $H_2O$ →$Ga_2O_3$ + $H_2$ + $C_2H_5$ ($\Delta G_{rxn}$ = -57 meV/atom)

Using these $\Delta G_{rxn}$, we calculated the relative nucleation rates shown in Figure S9. Enumerated reaction pathways are summarized in Table S5.

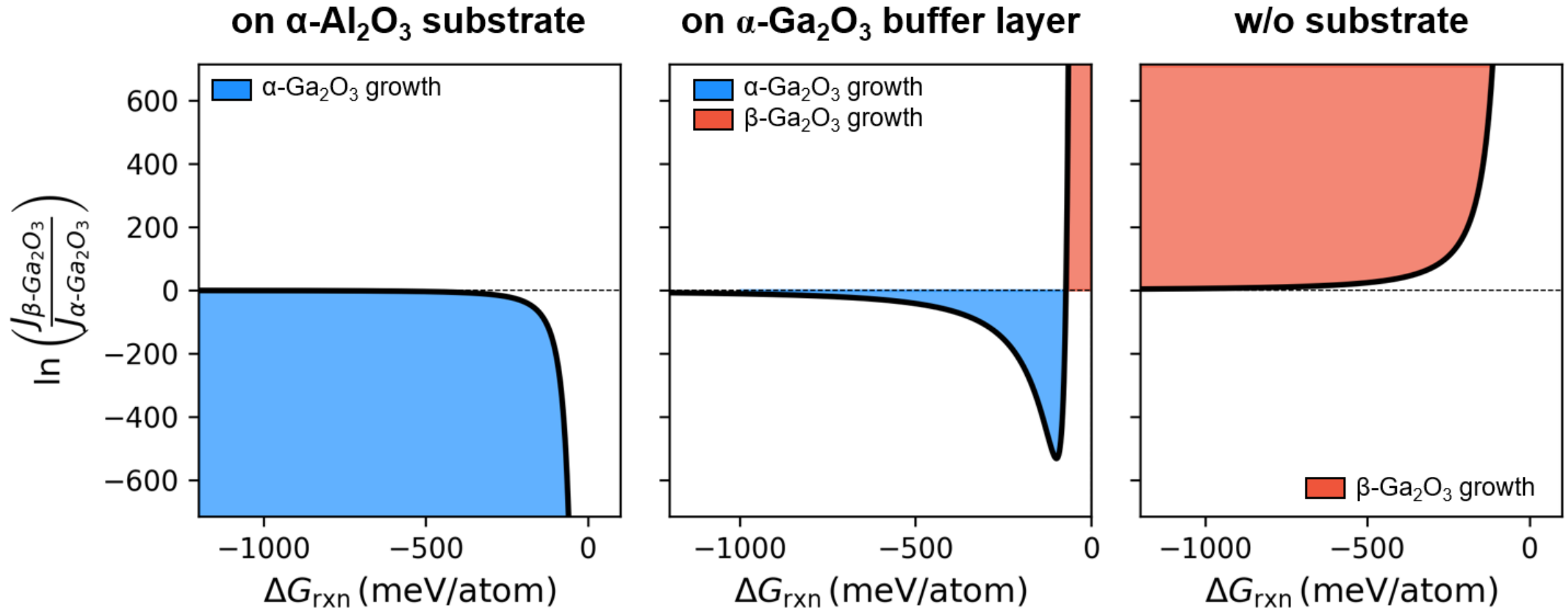

**Figure S7.** Calculated relative nucleation rates for α- and β-$Ga_2O_3$ heteroepitaxy across the accessible nucleation configurations.

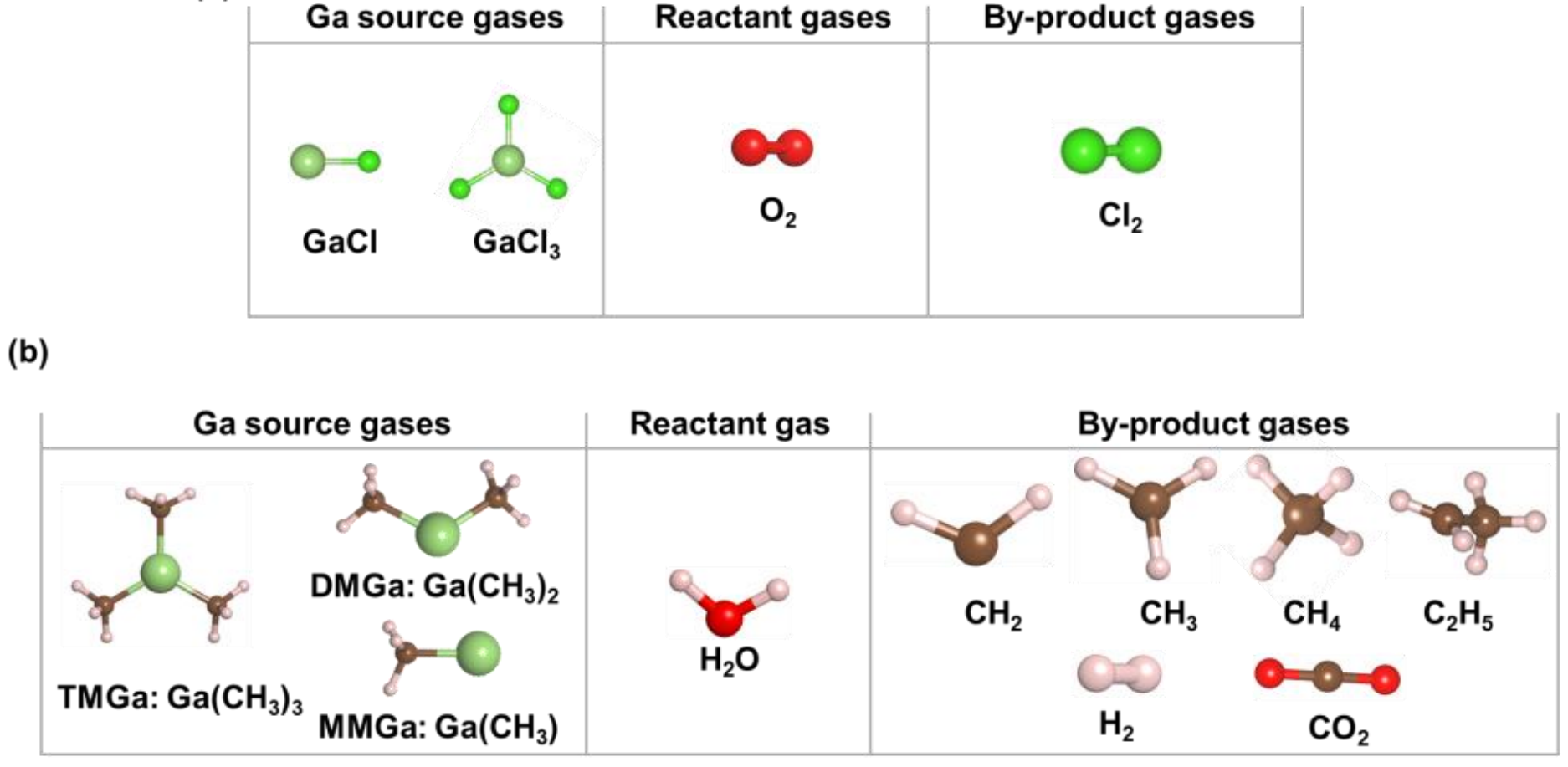

**Figure S8.** Possible reaction components in (a) β- and (b) α-$Ga_2O_3$ heteroepitaxy Source gases, reactant gases, and by-product gases in $Ga_2O_3$ reactions using (a) GaCl and (b) TMGa precursors.

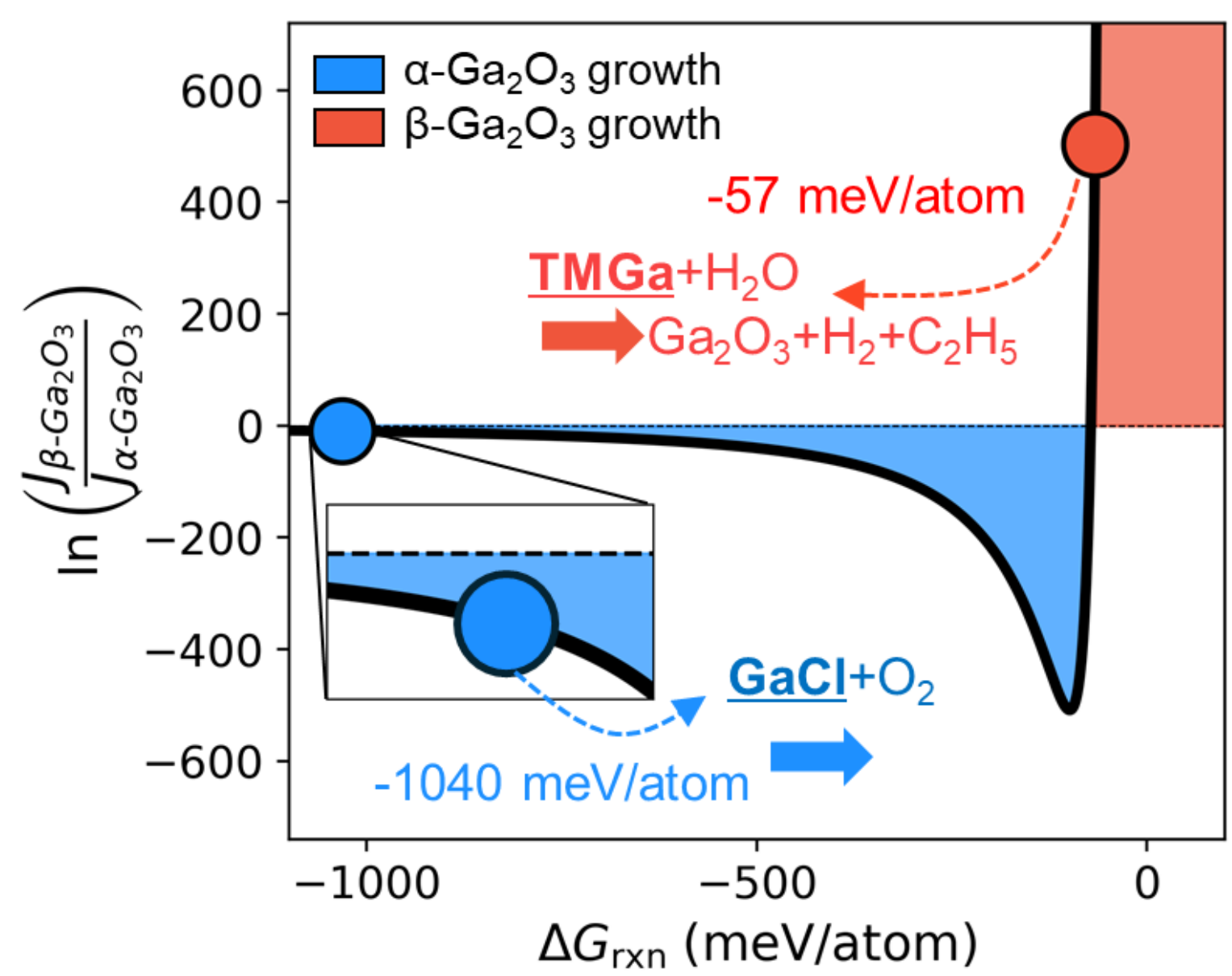

**Figure S9.** Calculated relative nucleation rates for the α- and β-$Ga_2O_3$ heteroepitaxy with reaction pathways.

| Reaction | Precursor | Flow of precursor | Reactant | Flow of reactant | Temperature (℃) | By-products | ΔG_rxn (meV/atom) | rxn |
|---|---|---|---|---|---|---|---|---|
| 1 | GaCl | 138 sccm | O2 | 80 sccm | 650 | GaCl3 | -1165.15 | GaCl + 0.5 O2 -> 0.3333 Ga2O3 + 0.3333 GaCl3 |
| 2 | GaCl | 138 sccm | O2 | 80 sccm | 650 | Cl2 | -1031.94 | GaCl + 0.75 O2 -> 0.5 Ga2O3 + 0.5 Cl2 |
| 3 | TMGa | 5 sccm | H2O | 400 sccm | 800 | H2, Ga(H3C)3 | -163.5 | 3 H2O + 3 GaH3C -> Ga2O3 + 3 H2 + Ga(H3C)3 |
| 4 | TMGa | 5 sccm | H2O | 400 sccm | 800 | H2, Ga(H3C)3 | -142.18 | 6 Ga(H3C)2 + 3 H2O -> Ga2O3 + 3 H2 + 4 Ga(H3C)3 |
| 5 | TMGa | 5 sccm | H2O | 400 sccm | 800 | H2, H2C | -84.41 | 3 H2O + 2 GaH3C -> Ga2O3 + 4 H2 + 2 H2C |
| 6 | TMGa | 5 sccm | H2O | 400 sccm | 800 | H2, Ga(H3C)2 | -77.46 | 3 H2O + 4 GaH3C -> Ga2O3 + 3 H2 + 2 Ga(H3C)2 |
| 7 | GaCl3 | 138 sccm | O2 | 80 sccm | 650 | Cl2 | -63.45 | 2 GaCl3 + 1.5 O2 -> Ga2O3 + 3 Cl2 |
| 8 | TMGa | 5 sccm | H2O | 400 sccm | 800 | H2, H5C2 | -57.4 | 3 H2O + 2 GaH3C -> Ga2O3 + 3.5 H2 + H5C2 |
| 9 | TMGa | 5 sccm | H2O | 400 sccm | 800 | H2, H2C | -35.59 | 2 Ga(H3C)2 + 3 H2O -> Ga2O3 + 5 H2 + 4 H2C |
| 10 | TMGa | 5 sccm | H2O | 400 sccm | 800 | H2, H5C2 | 2.42 | 2 Ga(H3C)2 + 3 H2O -> Ga2O3 + 4 H2 + 2 H5C2 |
| 11 | TMGa | 5 sccm | H2O | 400 sccm | 800 | H2, H3C | 43.8 | 3 H2O + 2 GaH3C -> Ga2O3 + 3 H2 + 2 H3C |
| 12 | TMGa | 5 sccm | H2O | 400 sccm | 800 | H2, H2C | 86.77 | 2 Ga(H3C)3 + 3 H2O -> Ga2O3 + 6 H2 + 6 H2C |
| 13 | TMGa | 5 sccm | H2O | 400 sccm | 800 | H2, H5C2 | 130.75 | 2 Ga(H3C)3 + 3 H2O -> Ga2O3 + 4.5 H2 + 3 H5C2 |
| 14 | TMGa | 5 sccm | H2O | 400 sccm | 800 | H2, H3C | 144.84 | 2 Ga(H3C)2 + 3 H2O -> Ga2O3 + 3 H2 + 4 H3C |
| 15 | TMGa | 5 sccm | H2O | 400 sccm | 800 | H2, H3C | 295.56 | 2 Ga(H3C)3 + 3 H2O -> Ga2O3 + 3 H2 + 6 H3C |

| 16 | TMGa | 5 sccm | H2O | 400 sccm | 800 | H2, H4C | 330.68 | 3 H2O + 2 GaH3C -> Ga2O3 + 2 H2 + 2 H4C |
|---|---|---|---|---|---|---|---|---|
| 17 | TMGa | 5 sccm | H2O | 400 sccm | 800 | H4C, GaH3C | 518.71 | 4 Ga(H3C)2 + 3 H2O -> Ga2O3 + 6 H4C + 2 GaH3C |
| 18 | TMGa | 5 sccm | H2O | 400 sccm | 800 | Ga, H4C | 527.45 | 3 H2O + 6 GaH3C -> Ga2O3 + 4 Ga + 6 H4C |
| 19 | TMGa | 5 sccm | H2O | 400 sccm | 800 | H2, H4C | 548.61 | 2 Ga(H3C)2 + 3 H2O -> Ga2O3 + H2 + 4 H4C |
| 20 | TMGa | 5 sccm | H2O | 400 sccm | 800 | Ga, H4C | 629.14 | 3 Ga(H3C)2 + 3 H2O -> Ga2O3 + Ga + 6 H4C |
| 21 | TMGa | 5 sccm | H2O | 400 sccm | 800 | H4C | 762.77 | 2 Ga(H3C)3 + 3 H2O -> Ga2O3 + 6 H4C |
| 22 | TMGa | 5 sccm | H2O | 400 sccm | 800 | H2, CO2 | 797.33 | 7 H2O + 2 GaH3C -> Ga2O3 + 10 H2 + 2 CO2 |
| 23 | TMGa | 5 sccm | H2O | 400 sccm | 800 | H2, CO2 | 1013.35 | 2 Ga(H3C)2 + 11 H2O -> Ga2O3 + 17 H2 + 4 CO2 |
| 24 | TMGa | 5 sccm | H2O | 400 sccm | 800 | H2, CO2 | 1154.92 | 2 Ga(H3C)3 + 15 H2O -> Ga2O3 + 24 H2 + 6 CO2 |

**Table S5**. Enumerated reaction list using GaCl and TMGa precursors.

## S8. Prediction of heteroepitaxial κ- and β-$Ga_2O_3$ nucleation on α-$Ga_2O_3$ buffer layer

By inputting Table S4 to S1.5, we calculated the relative nucleation rates shown in Figure S10. Based on experimental reports [10], we found the following reactions:

| Process | Precursor | Precursor flow rate | Reactant | Reactant flow rate | Growth temperature | Chamber pressure | Substrate |
|---|---|---|---|---|---|---|---|
| MOCVD | TEGa | [O2]/ [TEGa] =350 | O2 | 400 SCCM | 610 ℃ | 3~400 mbar | α-Al2O3 (001) |

1. $Ga(C_2H_5)_3 + O_2 \rightarrow Ga_2O_3 + H_2 + CH_2$, $P_{TEGa} = 0.4 \times 10^{-6}$ atm ($\Delta G_{rxn}$ = -33 meV/atom)
2. $Ga(C_2H_5)_3 + O_2 \rightarrow Ga_2O_3 + H_2 + CH_2$, $P_{TEGa} = 2.3 \times 10^{-6}$ atm ($\Delta G_{rxn}$ = -42 meV/atom)
3. $Ga(C_2H_5)_3 + O_2 \rightarrow Ga_2O_3 + H_2 + CH_2$, $P_{TEGa} = 4.0 \times 10^{-6}$ atm ($\Delta G_{rxn}$ = -45 meV/atom)
4. $Ga(C_2H_5)_3 + O_2 \rightarrow Ga_2O_3 + H_2 + CH_2$, $P_{TEGa} = 5.8 \times 10^{-6}$ atm ($\Delta G_{rxn}$ = -47 meV/atom)
5. $Ga(C_2H_5)_3 + O_2 \rightarrow Ga_2O_3 + H_2 + CH_2$, $P_{TEGa} = 46 \times 10^{-6}$ atm ($\Delta G_{rxn}$ = -57 meV/atom)

Using these $\Delta G_{rxn}$, we calculated the relative nucleation rates shown in Figure S11.

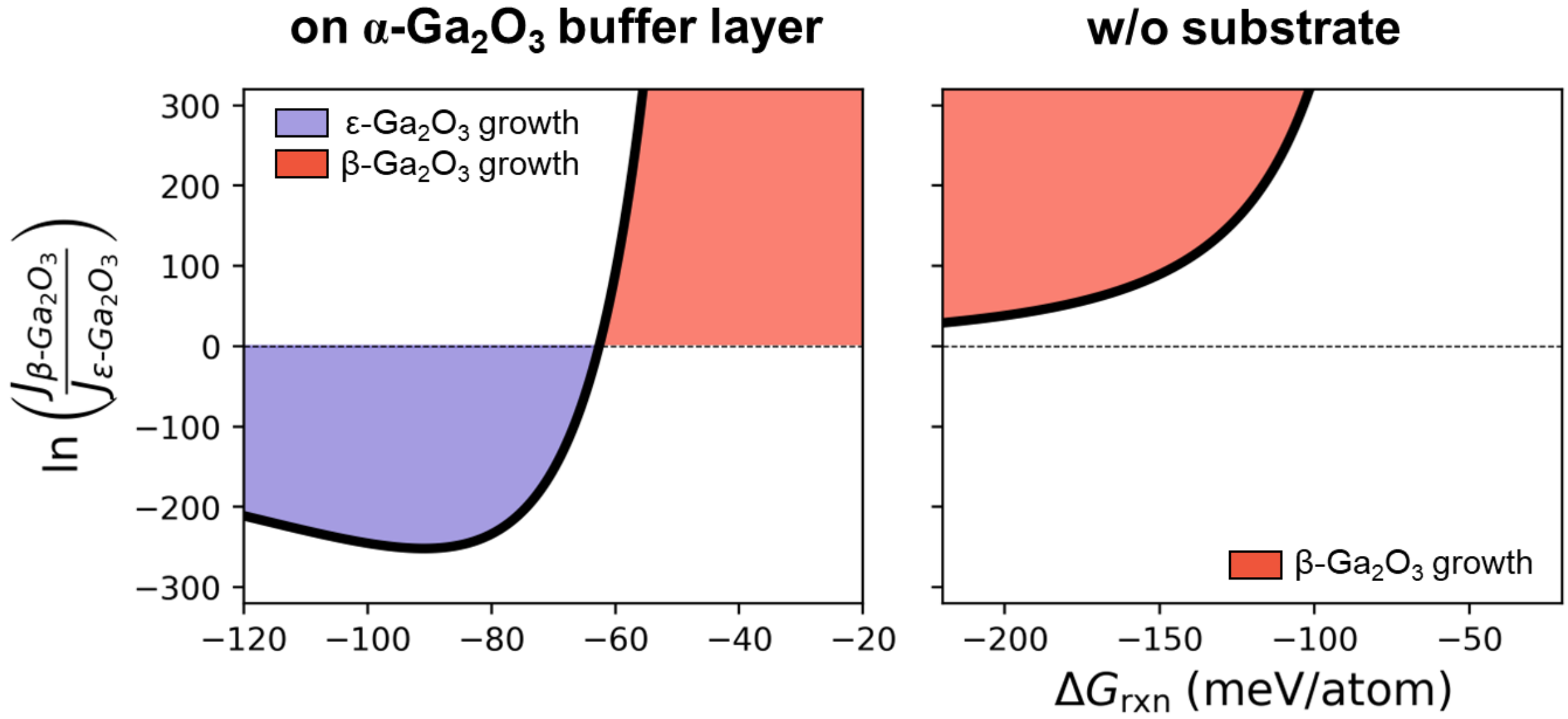

**Figure S10.** Calculated relative nucleation rates for the κ- and β-$Ga_2O_3$ heteroepitaxy

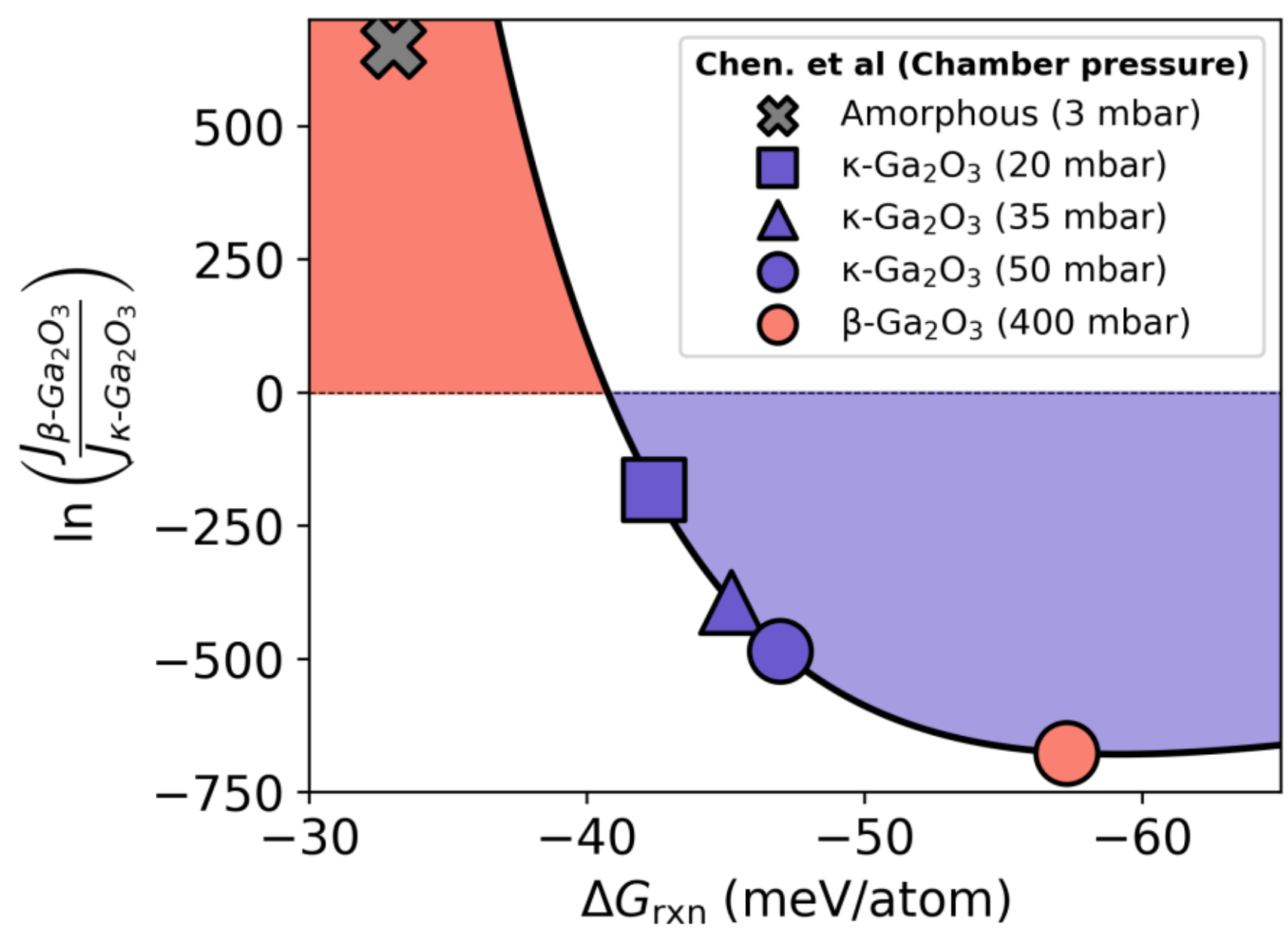

**Figure S11.** Calculated relative nucleation rates for the κ- and β-$Ga_2O_3$ heteroepitaxy with reaction pathways.

## S9. Nucleation-based T-P diagram of κ- and β-$Ga_2O_3$ on α-$Ga_2O_3$ buffer layer

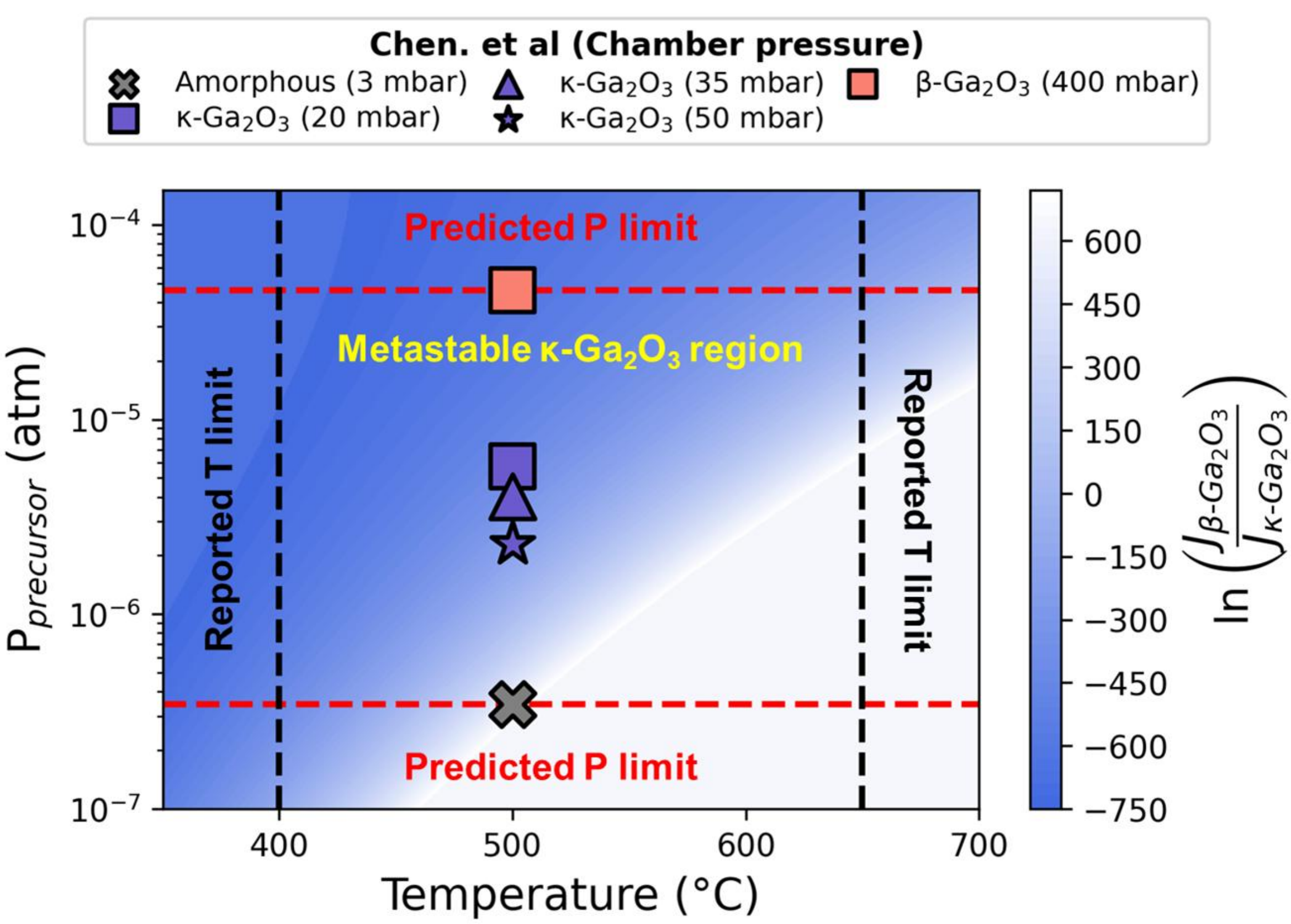

**Figure S12**. T-P diagram based experimental reports and relative nucleation rates between κ- and β-$Ga_2O_3$ on α-$Ga_2O_3$ buffer layer [10].

## S10. Building a nucleation-based phase diagram for polymorphic $Ga_2O_3$ deposition.

The nucleation-based phase diagram is identified by linking the relative nucleation rates of metastable α- and κ-phases to that of β-$Ga_2O_3$. Based on the observation that the α/κ-to-β nucleation rate trends are similar (Figure S13), we connect the two curves centered around the direction of the most favorable reaction energy, as shown in Figure 4a. Using process-derived reaction energies, the relative nucleation rates are pinpointed for β-$Ga_2O_3$ , κ-$Ga_2O_3$, α-$Ga_2O_3$, and the amorphous state. The process conditions extracted from the experimental literature [7,10], along with the corresponding reaction energies, were used as follows:

**Amorphous $Ga_2O_3$**: 450℃, $F_{GaCl3}$ = 4600 SCCM, $F_{O2}$=600 SCCM, $P_{GaCl3}$=0.884615, $P_{O2}$=0.115385, $\Delta G_{rxn}$=−68 meV/atom

**Heteroepitaxial α-$Ga_2O_3$**: 500℃, $F_{GaCl}$=3800 SCCM, $F_{O2}$=600 SCCM, $P_{Gacl}$=0.866667, $P_{O2}$=0.133333, $\Delta G_{rxn}$=−1040 meV/atom

**Heteroepitaxial κ-$Ga_2O_3$**: 650℃, $F_{GaCl}$=300 SCCM, $F_{O2}$= 800 SCCM, $P_{GaCl}$=0.197368, $P_{O2}$=0.526316, $\Delta G_{rxn}$=−1024 meV/atom

**Heteroepitaxial β-$Ga_2O_3$**: 850℃, $F_{TMGa}$=85 SCCM, $F_{O2}$=2800 SCCM, $P_{TMGa}$=0.001745, $P_{O2}$=0.057466, $\Delta G_{rxn}$=−121 meV/atom

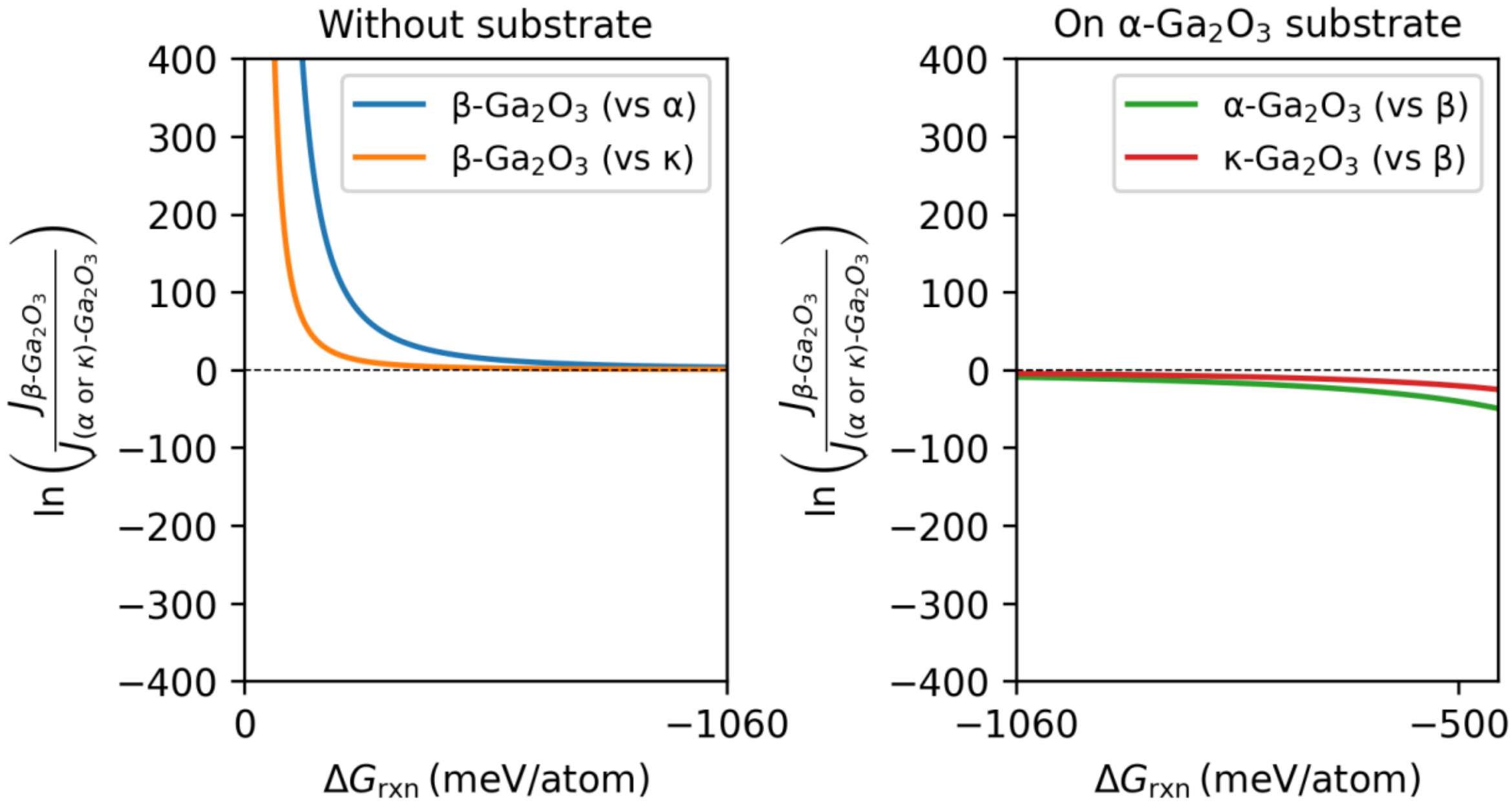

**Figure S13**. Calculated relative nucleation rates between β-$Ga_2O_3$ and α- and κ- phases on (a) without substrate, and (b) On α-$Ga_2O_3$ buffer layer.

## S11. Prediction of heteroepitaxial $TiO_2$ nucleation on $SrTiO_3$ substrate

From experimental reports [20, 21], we identified the information on $TiO_2$ deposition as follows:

Anatase conditions: 650°C, $F_{TDMAT}$ = 4×10$^{-7}$ mol/min, $F_{O2}$ = 70 SCCM, $P_{chamber}$ = 10 torr, substrate = $SrTiO_3$

Rutile conditions: 1227°C, $F_{TTIP}$ = 0.4 kpa, $F_{O2}$ = 0.4 kpa, $P_{chamber}$ = 0.4 kpa, substrate = $SrTiO_3$

We calculated and collected the polymorphic energetics of $TiO_2$, as shown in Table S6. By inputting Table S6 to S1.5, we calculated the relative nucleation rates shown in Figure S15. We then configured the set of plausible reaction components (Figure S16) and screened the following reactions:

| Name | Value | Unit |
|---|---|---|
| $\gamma_{surf,a-TiO_2\,(001)}$ | 67.5[1] | meV/Å$^2$ |
| $\gamma_{surf,r-TiO_2\,(111)}$ | 60[2] | meV/Å$^2$ |
| $\gamma_{surf,a-TiO_2\,(101)}$ | 58[3] | meV/Å$^2$ |
| $\gamma_{surf,r-TiO_2\,(110)}$ | 53[3] | meV/Å$^2$ |
| $\gamma_{inter,a-TiO_2\,(001)}^{epi,SrTiO_3\,(001)}$ | 15.9[4] | meV/Å$^2$ |
| $\gamma_{inter,r-TiO_2\,(111)}^{epi,SrTiO_3\,(001)}$ | 62.4[4] | meV/Å$^2$ |
| $n_{a-TiO_2}$ | 0.083 | atom/Å$^3$ |
| $n_{r-TiO_2}$ | 0.0956 | atom/Å$^3$ |
| $\Delta E_{epi}^{SrTiO_3}$ | 60[4] | meV/atom |
| $\Delta E_{helm}$ | 5 | meV/atom |

**Table S6.** Determined parameters to calculate nucleation rates for anatase and rutile-$TiO_2$ competition.

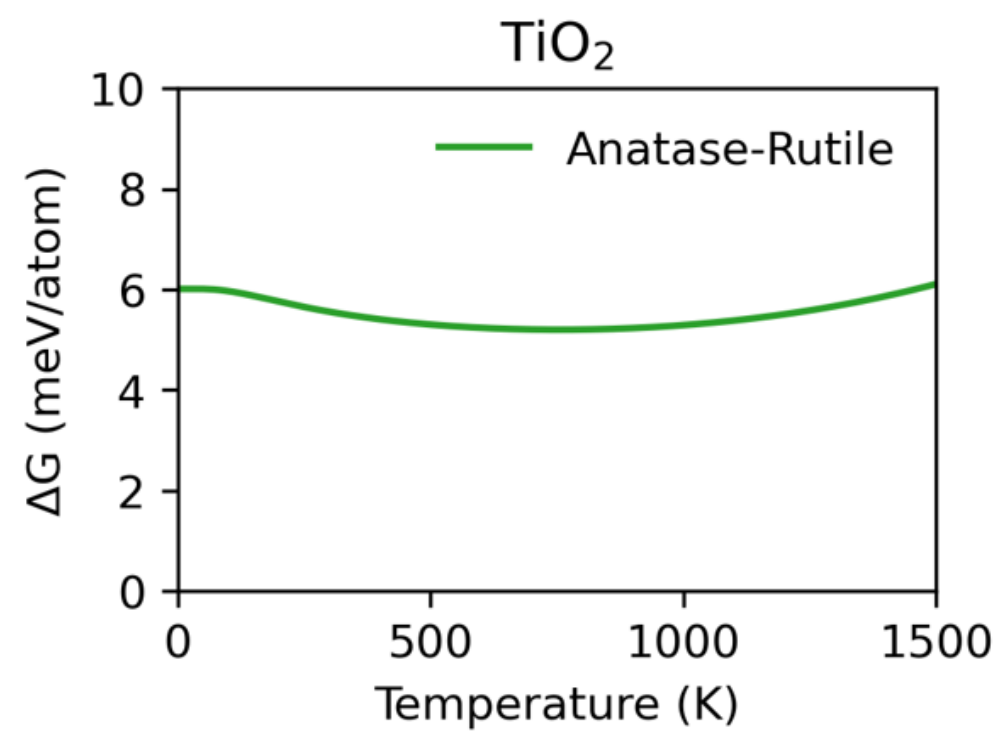

**Figure S14.** Calculated finite-temperature Gibbs free energy for anatase and rutile $TiO_2$.

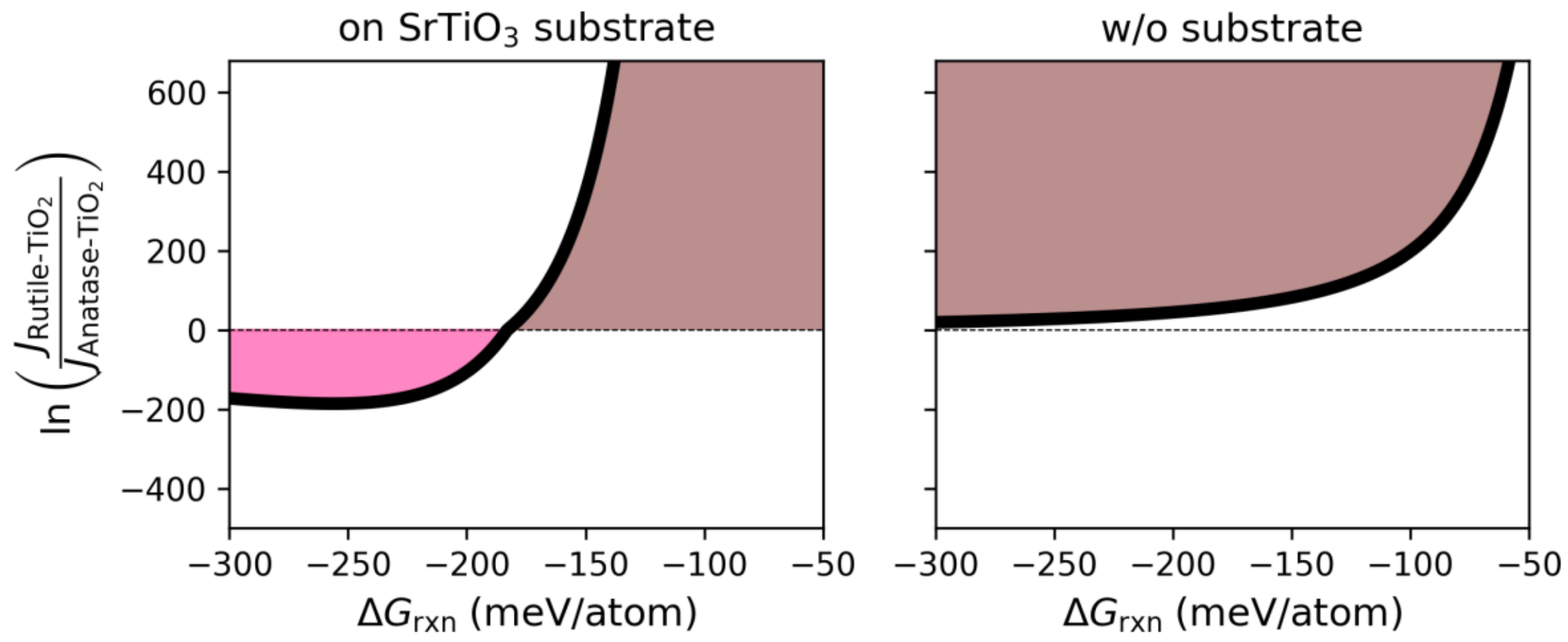

**Figure S15.** Calculated relative nucleation rates for the anatase and rutile $TiO_2$

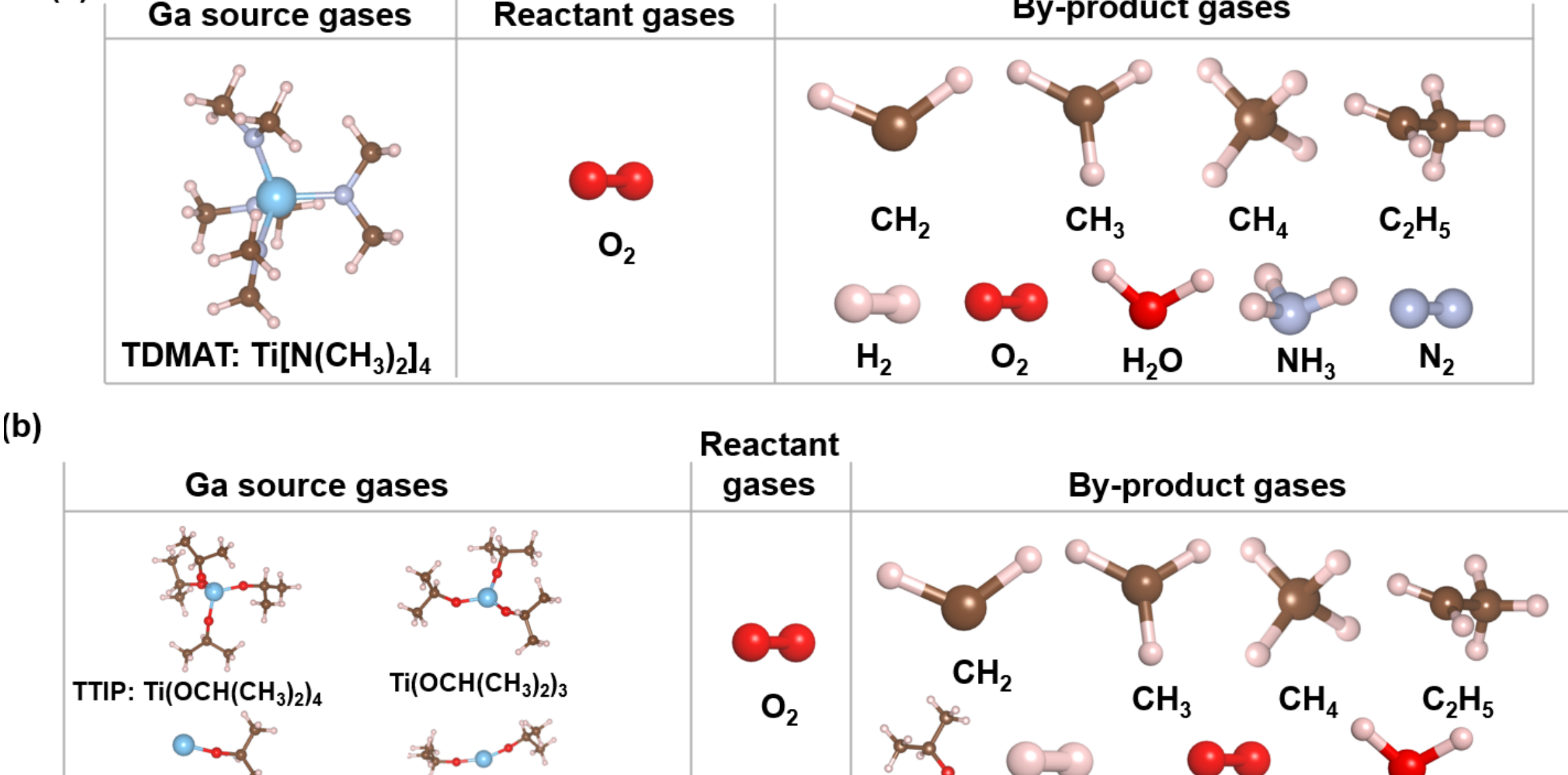

**Figure S16.** Source gases, reactant gases, and by-product gases in $TiO_2$ reactions using (a) TDMAT and (b) TTIP precursors.